\documentclass{aa}  

\usepackage{graphicx}
\usepackage{txfonts}
\usepackage[]{hyperref}
\def\arcsec{\hbox{$^{\prime\prime}$}}

\def\farcs{\hbox{$.\!\!^{\prime\prime}$}}

\begin{document}

\title{The dynamical state of the globular clusters Rup~106 and IC\,4499}

   \author{G. Beccari
          \inst{1}
          \and
          M. Cadelano\inst{2,3}
          \and
          E. Dalessandro\inst{3}
          }

   \institute{European Southern Observatory, Karl-Schwarzschild-Strasse 2, 85748  Garching bei M\"unchen, Germany\\
              \email{gbeccari@eso.org}
              \and
              Dipartimento di Fisica e Astronomia ``Augusto Righi'', Universit\'a degli Studi di Bologna, Via Gobetti 93/2, I-40129 Bologna, Italy 
         \and
             INAF-Astrophysics and Space Science Observatory Bologna, Via Gobetti 93/3, I-40129 Bologna, Italy
              }

   \date{Received September 15, 1996; accepted March 16, 1997}

 
  \abstract {The dynamical evolution of globular clusters is theoretically described by a series of well known events typical of N-body systems. Still, the identification of observational signatures able to empirically describe the stage of dynamical evolution of a stellar system of the density typical of a globular cluster, represents a challenge. In this paper we study the dynamical age of the globular clusters Rup~106 and IC~4499. To this aim, we study the radial distribution of the Blue Straggler Stars via the A$^+$ parameter and of the slope of the Main Sequence Mass Function. Both tracers show that Rup 106 and IC 4499 are dynamically young clusters where dynamical friction has just started to segregate massive stars towards the clusters' centre. In fact, we observe that the Blue Straggler stars are more centrally concentrated in both clusters than the reference population. On the same line, we find that in both cases the slope of the mass function significantly decreases as a function of the cluster-centric distances. This result provides additional support for the use of the the radial distribution of the blue stragglers as a powerful observationally convenient indicator of the cluster dynamical age. } 


   \maketitle
%

\section{Introduction}
\label{sec_int}
Globular clusters (GCs) are the oldest ($\sim12$ Gyr), most massive ($10^4 - 10^6$ M$_{\odot}$) and dense stellar aggregates in the Galaxy. They are populated by millions of stars, whose age, distance, and chemical composition can be determined with high accuracy. For this reason,
GCs play a crucial role in the current understanding of stellar
and dynamical evolution.
During their long life, GCs experience all the most relevant dynamical events foreseen in a {\it N}-body system such as: gravothermal instability, violent relaxation, energy equipartition, 2-body and higher order collisions, binary formation and heating, ``erosion'' by the tidal interaction with external fields~\citep{Meylan1997}. 
The internal dynamics of such massive stellar aggregates affect objects of any mass and it can be efficiently probed by means of massive test particles, like blue straggler stars (BSSs), binaries and
millisecond pulsars~\citep[e.g.][]{Bailyn1995,Beccari2006,prager17,mapelli2018,sing2019}. Among
them, BSSs have been largely used for this purpose, since
they are numerous and relatively easy to detect (e.g., \citealt{Ferraro2012, f18}).

BSSs are the by-product of the complex interplay between cluster dynamics and stellar evolution. They are defined as stars  hotter (and therefore bluer) and brighter than Main Sequence Turnoff (MSTO) stars within a cluster. 
BSSs  form either through mass-transfer \citep{Leonard1996} or collision events \citep{Lombardi1995}. 
Both formation channels are expected to work simultaneously within a given cluster \citep{Knigge2009,ferraro2009,Dalessandro2013}. However, the relative importance of each process
likely depends on the physical properties of the parent cluster
~\citep[e.g.][]{Sollima2008,leigh2013,sills2013}.
Independently on their formation mechanism, BSSs are a population of heavy
objects ($\sim1-1.6 M_{\odot}$) orbiting in a sea of 3-4 times less massive stars.
For this reason, BSSs can be used
as powerful gravitational probe particles to investigate key
dynamical processes (such as mass segregation) characterising
the dynamical evolution of star clusters.
Indeed in a series of paper \citep{Ferraro2012,Lanzoni2016,f18} our group has shown that the morphology of the BSS
radial distribution is shaped
by the action of dynamical friction (DF), which drives the objects
more massive than the average toward the cluster centre, with an
efficiency that decreases with increasing radial distance. 

Driven by the same physical processes, the radial variation of the luminosity or mass functions
(LF; MF) of MS stars can provide information about the effect of cluster internal
dynamics on stars in a wide range of masses, including the
faint-end of the MS where most of the cluster mass lies. In
relaxed stellar systems, the slope of the MFs is expected to vary
as a function of the distance from the cluster centre, with indexes decreasing as the distance increases, because of the
differential effect of mass segregation~\citep{webb17, ca20}.



Here we present results about the dynamical state of the GCs IC\,4499 and Ruprecht\,106 (Rup~106 from now on). In our study we use photometric catalogues obtained combining Hubble Space Telescope (HST) and ground-based wide field data. The vast data set allows us to resolve the stellar population of the clusters from the very centre to the their outer regions.

IC\,4499 is a low-density ($c=1.21$) cluster in the southern hemisphere at RA=15h00m18.45s, DEC=-82h12m49.3s lying in the intermediate-outer Galactic halo at a distance of 15.7\,kpc from the Galactic centre \citep[][2010 edition]{Harris2010}. 
It aroused attention in the past due to its high frequency of variable stars~\citep{Walker1996}, a large number of BSS~\citep{Dotter2011} and for having being suggested as one of the very few GC not showing evidence of multiple populations with light-element abundance variations (but see \citealt{dalessandro18}). 
\citet{Ferraro1995} suggested an unusual 3 to 4 Gyr younger age for this cluster compared to clusters with similar metallicity. This result was contradicted by \citet{Dotter2011} suggesting an age of 12.0\,$\pm$\,0.75\,Gyr and [Fe/H]\,$\sim$\,-1.6.

Rup~106 is a low mass~\citep[$3.4\times10^4$M$_{\odot}$][]{Baumgardt19}, metal poor GC lying about 25\,kpc from the galactic centre, first examined by \citet{Buonanno1990,Buonanno1993}. Similarly to IC 4499, Rup 106 was suggested~\citep{Buonanno1993} to be 4-5 Gyr younger than GCs with similar metallicty. 
More recently, \citet{Dotter2011} suggested an age of 11.5\,Gyrs with [Fe/H]\,$\sim$\,-1.5
while~\citet[][]{Villanova2013} demonstrated that Rup~106 is one of the few GC which convincingly does not show evidences of the presence of multiple stellar populations~\citep[see also][]{fuj2021}.

Moreover \citet{FusiPecci1995} suggests for both IC\,4499 and Rup~106 a possible membership to a tidal stream around the Milky Way which could suggest a preceding capture from an outer galactic source. Recently ~\citet[][]{massari19} using Gaia DR2 data found that IC~4499 is likely associated to the Sequoia galaxy merging event while similarly Rup~106 is associated to the progenitor of the Helmi stream.

In Section \ref{sect_ObsData} we present the observation and data reduction process used. We derive the cluster' structural parameters in Section~\ref{sec_param} while in Section \ref{sec_bss} use the radial distribution of the BSS of the clusters to investigate their dynamical state. The later is also studied looking at the signature of mass segregation imprinted in the radial distribution of the slope of the Mass Function (Section~\ref{sec_mf}).

\section{Observations and data reduction}
\label{sect_ObsData}
The main scope of this paper is to empirically investigate the dynamical
state of the GCs Rup~106 and IC\,4499 using the radial distribution of BSSs and any radial change of the slope of the MF ($\alpha$). Both approaches require a data-set able to sample the stellar population in both clusters over a comparable physical area.

To this aim, we used a combination of ground based wide field images to sample the brightest populations of the clusters from the central region to the clusters' tidal radii (r$_t$) and deep high resolution images acquired with the Advanced Camera for Survey (ACS) on board of the Hubble Space Telescope (HST) to sample the MS stars to the very low mass regime out to the cluster's half-mass radii (r$_h$).

\begin{figure*}[ht!]
 \scalebox{0.3}{\includegraphics{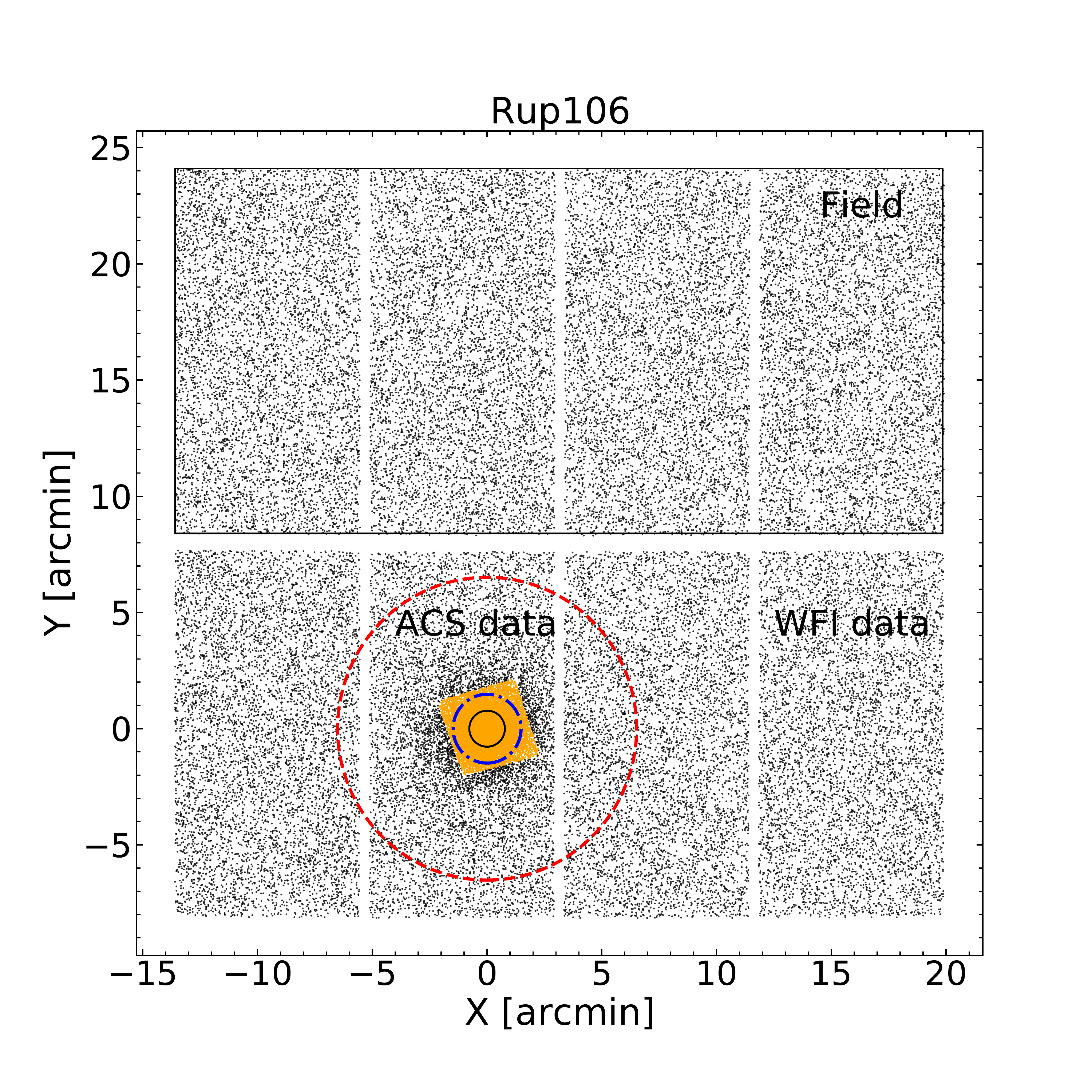}}
 \scalebox{0.3}{\includegraphics{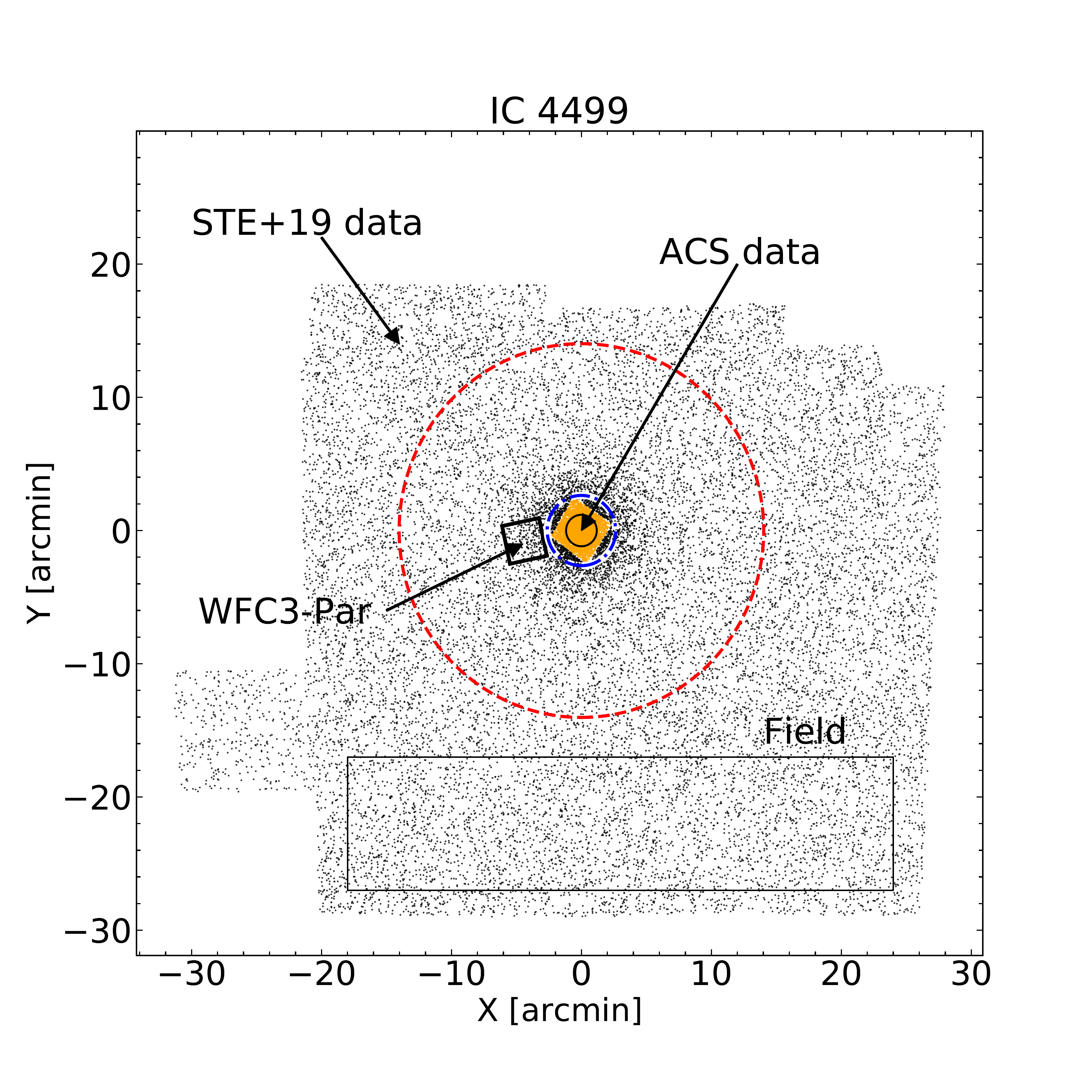}}
  \caption{Data FoV. Composite FOVs showing the stars sampled in Rup 106 (left panel) and IC 4499 (right panel). The central field identifies the stars sampled with the ACS catalogues (orange points). The solid (black), dot-dashed (blue) and dashed (red) circles mark the core, half-mass and tidal radii of the clusters, respectively. The area used to estimate the field contamination and the one covered by the WFC3 parallel observations are also labelled.}
  \label{fig_maps}
\end{figure*}

Rup~106 and IC~4499 were both observed with the ACS under the HST Proposal 11586 (PI Dotter).  The clusters were imaged with the cluster's core positioned at the centre of the ACS Field of View (FoV), in the F606W (V band) and F814W (I band) filters for total exposure times of 2250 s and 2340 s (Rup~106), and 2412 s and 2544 s (IC\,4499), respectively. A dithering pattern was adopted in both cases to fill the gaps in ACS CCD mosaic. The ACS data alone allow us to sample the MS stars of Rup~106 out to the cluster's r$_h$=88\farcs{83} (dash-dotted circle in Fig.~\ref{fig_maps}; see also Sec.~\ref{sec_param}). 

IC~4499 has a r$_h=158\farcs{15}$ which falls out of the ACS FoV (see Fig.~\ref{fig_maps}). In this case we also retrieved from the HST archive a set of images taken with the Wide Field Camera 3 (WFC3) as part of HST parallel observations. In particular, $2\times600$s and $2\times800$s parallel observations
with the F606W and F814W filters were obtained with the WFC3 via the HST Proposal 14235 (PI Sohn) to observe the external region of IC~4499. These parallel observations allow us to sample the MS stars in IC~4499 outside of the core region and beyond the r$_h$, as in the case of Rup~106.

In order to sample the brightest stars in Rup~106, we complemented the ACS sample with a catalogue obtained using a number wide-field images acquired under the program 71.D-0220(A) (PI: Ortolani) with the Wide Field Imager (WFI)
at the MPE2.2m telescope (La Silla) using the V89\_ESO843 and IcIwp\_ESO845 filters (hereafter V and I, respectively). Thanks to its wide FoV ($34'\times34'$) and resolution ($0\farcs234$), the WFI camera allows us to sample the brightest stellar population of the cluster in a region well beyond its r$_t$. Finally, we used the wide field catalogue of IC~4499 published by~\citep{ste19}, offering a complete star sample out to a radius of $842\arcsec$ or $14'$. The catalogue combines a sample of images from various telescopes and was previously used by~\citet{Walker2011} to study the formation history of IC~4499. A detailed description of the images adopted to create the photometric catalogue is available in table 1 of~\citet[][hereafter STE19]{ste19}. 

\begin{table}

 \centering
  \caption{Log of the observations}
  \begin{tabular}{@{}cccc@{}}
  \hline
   Data-set     & Number of exposures & Filter & Exposure time \\
            &                 &        &     (s) \\
 \hline
\multicolumn{4}{c}{Rup~106}\\
 \hline
 WFI  & 2 & $V$ & 200 \\
      & 2 & $I$ & 180 \\
  ACS   & 4 & $V_{606}$ & 550 \\
       & 4 & $I_{814}$ & 585 \\
\hline
\multicolumn{4}{c}{IC\,4499}\\
 \hline
  ACS   & 4 & $V_{606}$ & 603 \\
       & 4 & $I_{814}$ & 636 \\
  WFC3   & 2 & $V_{606}$ & 600 \\
       & 2 & $I_{814}$ & 800 \\
\label{tab_obs}
\end{tabular}

\end{table}

We give in table~\ref{tab_obs} a summary of the observations adopted in this work. We show in Fig.~\ref{fig_maps} the maps of the surveyed areas around  
Rup~106 (left panel) and IC~4499 (right panel). The location in the FoV of the different data-sets are
labelled in the figures. In both panels, the solid, dash-dotted and dashed circles indicate the position of the core radius ($r_c$), the half-mass radius ($r_h$) and tidal radius ($r_t$) of the clusters respectively, as estimated in this work (see Section \ref{sec_param}). 
We will use the ground based catalogues alone to calculate the radial density profiles of the clusters and estimate their structural parameters.  The ACS catalogues combined with the ground based catalogues will be used to study the radial distribution of the BSS in the two clusters (Section~\ref{sec_param}). To this purpose, the we will use the ground-based catalogues to sample the stellar population outside the FoV of the ACS data (see Fig.~\ref{fig_maps}). Finally, the ACS catalogues and the WFC3 parallel observations in the case of IC~4499, will be used to measure how the slope of the Mass Functions (MFs) changes as a function of the distance from the cluster's centres (Section~\ref{sec_mf}). 


\subsection{Data Reduction}
\label{sec_red}

We retrieved from the Mikulski Archive for Space Telescopes (MAST)~\footnote{https://archive.stsci.edu/} the ACS and WFC3 \texttt{flc} images, which are calibrated and corrected for Charge Transfer Efficiency (CTE) loss. 
We also
downloaded the raw WFI images together with the BIAS and FLAT-FIELD images from ESO archive~\footnote{http://archive.eso.org}. The FLAT-FIELD and BIAS correction was performed with the data reduction software Theli~\citep{theli05,theli13}.

We performed the data reduction of the HST (both ACS and WFC3) and WFI data-sets through an accurate modelling of the Point Spread Function (PSF) followed by a standard PSF-fitting procedure using \texttt{DAOPHOTII}~\citep[][]{ste87,ste94}. In short, we first 
used from 30 to 50 well-sampled and not saturated stars in each image to model the PSF.
We then used \texttt{DAOPHOTII/ALLFRAME} to perform the PSF fitting on the HST and WFI data-sets, separately. The magnitude of each star included in the final photometric catalogues is the average of at least 3 measures in each band, while we adopted the standard deviation as the associated photometric error. 

The instrumental V, I magnitudes of the ACS sample were calibrated into the Johnson standard system using the recipe in~\citet{Sirianni2005}. In short, a sample of bright isolated stars were used to transform the instrumental magnitudes to a fixed aperture of $0\farcs5$. Then, the extrapolation to infinite was performed by using the values listed in Table 5 of~\citet{Sirianni2005}. The magnitudes were finally transformed into the Johnson system using equation 12 from ~\citet{Sirianni2005}.


\begin{figure*}[ht]
  \scalebox{0.5}{\includegraphics{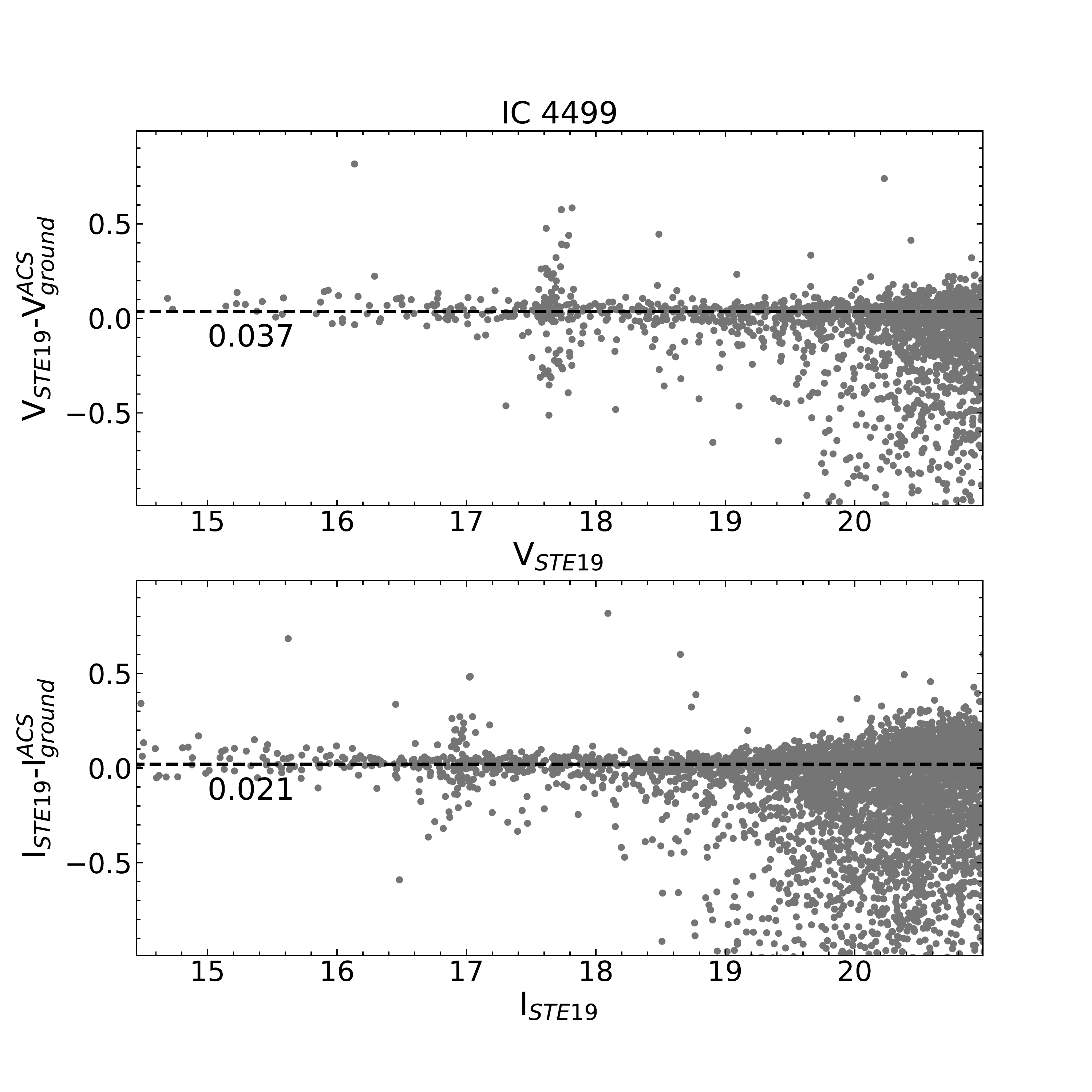}}
  \caption{Photometric calibration. Comparison between the V and I (upper and lower panels, respectively) photometry of IC~4499 calibrated into the Johnson photometric system for the stars in common between the ACS and the photometric catalogue of~\citep{ste19}. The horizontal line shows the best linear fit to the observed distribution.}
  \label{fig_cal}
\end{figure*}

For a sanity check we show in figure~\ref{fig_cal} the difference between the calibrated magnitudes of the stars in IC~4499 in common between the ACS and the STE19 catalogues. The plots demonstrate that using the calibration recipe described in~\citet{Sirianni2005} we achieve an excellent photometric homogeneity. Still, the best fit lines (dashed lines) indicate that there is a residual difference of 0.037 and 0.021 mag in V and I respectively, between the ACS and the STE19 photometry. Such a small residual offset is due to the difference between the Zeropoints from~\citep{Sirianni2005} and the latest implemented by the ACS Team~\footnote{https://www.stsci.edu/hst/instrumentation/acs/data-analysis/zeropoints}. We hence decided to apply such offsets to the ACS catalogue in order to obtain homogeneous V and I magnitudes between the ACS and the STE19 data-sets.
Similarly, adopting the Zero point available on the HST web pages\footnote{https://www.stsci.edu/hst/instrumentation/wfc3/data-analysis/photometric-calibration} we calibrated the instrumental magnitudes of the stars detected in the WFC3 data into VEGAMAG. Then, we used more than 100 stars in common with the STE19 catalogue, to calibrate the WFC3 catalogue against the STE19 photometry.


The calibration of the WFI catalogue of Rup~106 was done in 2 steps. First, we calibrated the V and I magnitudes of the WFI catalogue using 1174 stars in common with the UCAC4 Catalogue~\citep{zac12}. This first step allowed us to obtain an homogeneous calibration over the entire WFI FoV. We then transformed the magnitude of the stars of the WFI photometric catalogue into the ACS-Johnson system using more than 200 stars in common between the ACS and WFI catalogue. 

We obtained the absolute astrometric solution for the stars in both ACS catalogues and in the WFI one using several hundreds stars in common with the Gaia-DR2 catalogue~\citep{gaia1,gaia2}. The final rms global accuracy is of $0\farcs3$, in both right ascension ($\alpha$) and declination ($\delta$). 

\begin{figure*}[ht!]
 \scalebox{0.3}{\includegraphics{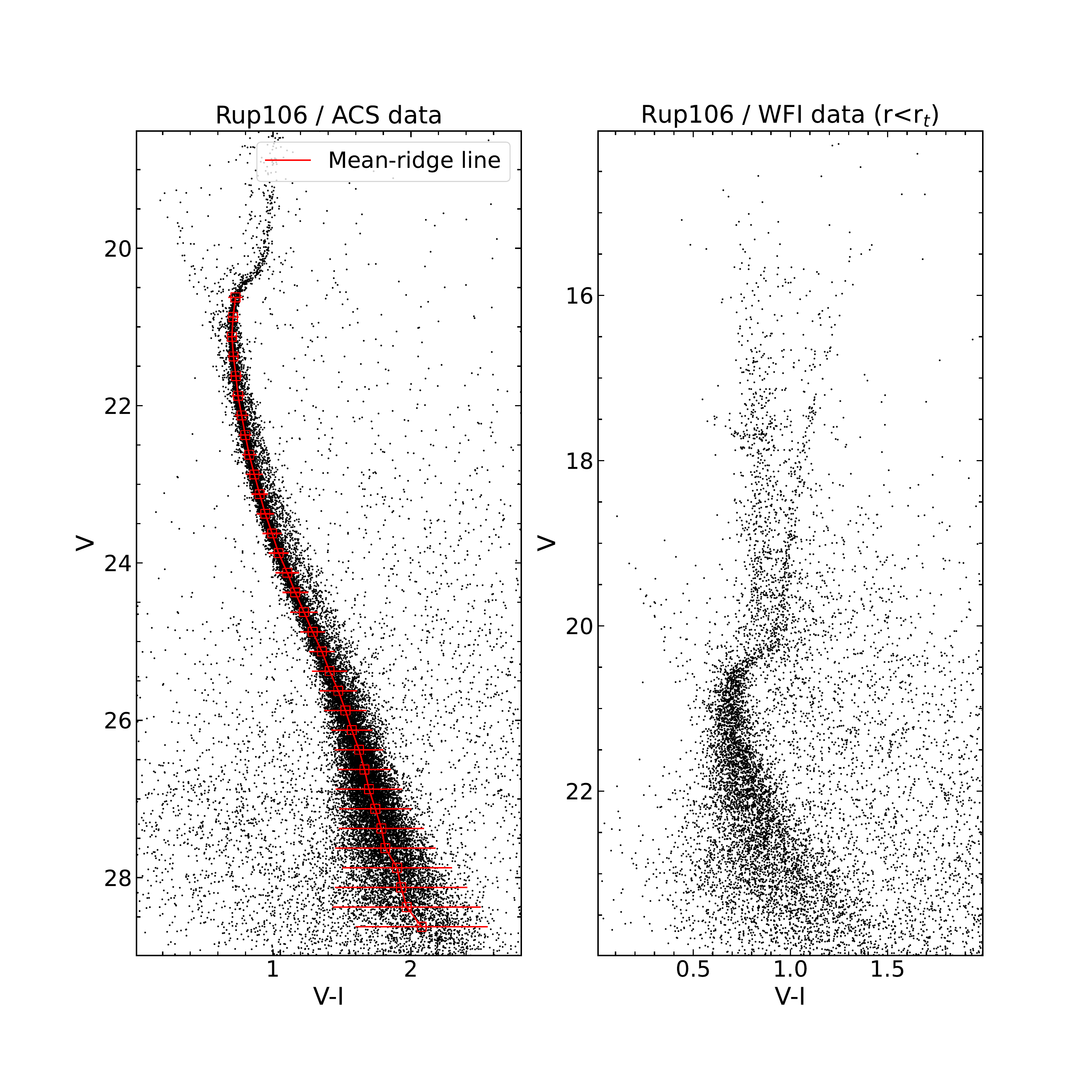}}
 \scalebox{0.3}{\includegraphics{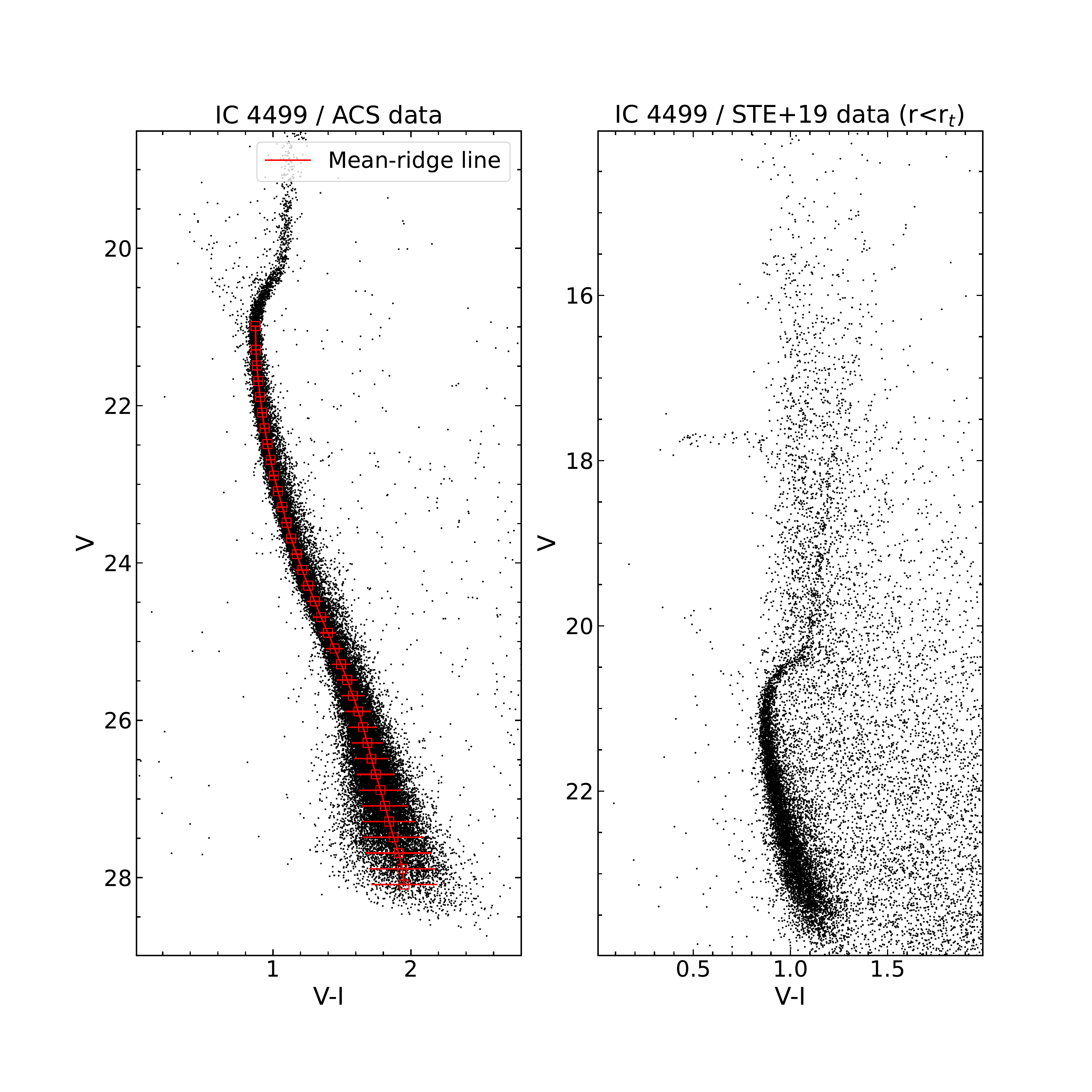}}
  \caption{Colour-magnitude diagrams of Rup 106 (two plots on the left-hand side) and IC 4499 (two plots on the right-hand side). The MS Mean-Ridge calculated on the HST data is shown as a red solid vertical line for both clusters.}
  \label{fig_cmd}
\end{figure*}


We show in Figure \ref{fig_cmd} the Colour-Magnitude-Diagrams (CMDs) of the cluster Rup~106 (two plots on the left-hand side) and IC~4499 (two plots on the right-hand side). We report at the top of each panel the name of the cluster and the adopted data-set while the MS Mean-Ridge calculated on the HST data is shown as a red solid vertical line for both clusters. Clearly, the ACS catalogues allows us to sample the clusters' stars from the population of BSS sampled in a region almost 2 magnitudes brighter with respect to the clusters' MSTO (V$_{TO}\sim21$), down to the very low mass end of the MS at V$\sim29$.  On the other side, the ground based catalogue sample the stars of the clusters below the MSTO and well beyond the clusters'$r_t$.

\subsection{Artificial Stars Experiment}
\label{sect_CompTest}

In this work we use detailed star counts to calculate the clusters' radial density profiles and derive the cluster's physical parameters (Sec~\ref{sec_param}). Also, we observationally investigate the dynamical status of Rup~106 and IC~4499 using the radial distribution of the BSS (Sec~\ref{sec_bss}) and of the slope of the clusters' MF (Sec.\ref{sec_mf}). It is hence
important to quantitatively assess the quality of the data in terms of completeness of the sample. The latter is defined as the amount of clusters' stars that we lost during the reduction analysis described in Sec.~\ref{sec_red}, as a function of the stellar magnitudes and radial position in the cluster. 

We study the photometric completeness following the recipe described in~\citet{Bellazzini2002} on the HST data-set which will be used to study the mass function of the clusters. As we explain later, we assess the completeness of the ground-base photometry from the HST catalogues.

First we created a catalogue of artificial stars whose V magnitude is extracted from the clusters' observed luminosity functions derived using the HST data. We artificially increase the number of stars towards the low luminosity/magnitude regimes. This is done in order to not over populate
the images with bright stars while at the same time well-sampling the low-magnitude regimes, where the statistic is crucial to
assess the fraction of stars lost during the data-reduction process. The I magnitude is assigned via the V-I colour obtained by interpolating each V magnitude using the
clusters MS mean-ridge line (see red solid lines in Fig.~\ref{fig_cmd}). Then, we add the stars on each V and  I HST
image adopting their specific PSF models by using the routine \texttt{DAOPHOTII/ADD}. 
Once the images with the artificial stars are created, we follow the same data-reduction strategy described
in Sec.~\ref{sec_red}. At the end of the data-reduction process, a star is considered recovered not only if
it was successfully measured in both magnitudes as for the real stars, but also if the difference between the assigned magnitude (input) 
and the one measured after the PSF fitting process (output) is $<0.75$ (both in V and I, independently). In fact, a star showing a larger difference in one of the magnitudes
is considered to be an unresolved blended source and hence lost~\citep{Bellazzini2002}. 
We stress here that stars are randomly placed on the FoV covered by the HST images, following a regular grid such that 
a minimum distance equal to $3\times$ the Full Width Half Maximum is
guaranteed between two simulated stars. This is critical in order to not introduce 
artificial stellar crowding to the images. The experiments
are repeated until a total of 100,000 stars is simulated 
in each ACS pointing and about 60,000 in the WFC3 parallel ones.


\begin{figure}
  \resizebox{0.85\hsize}{!}{\includegraphics{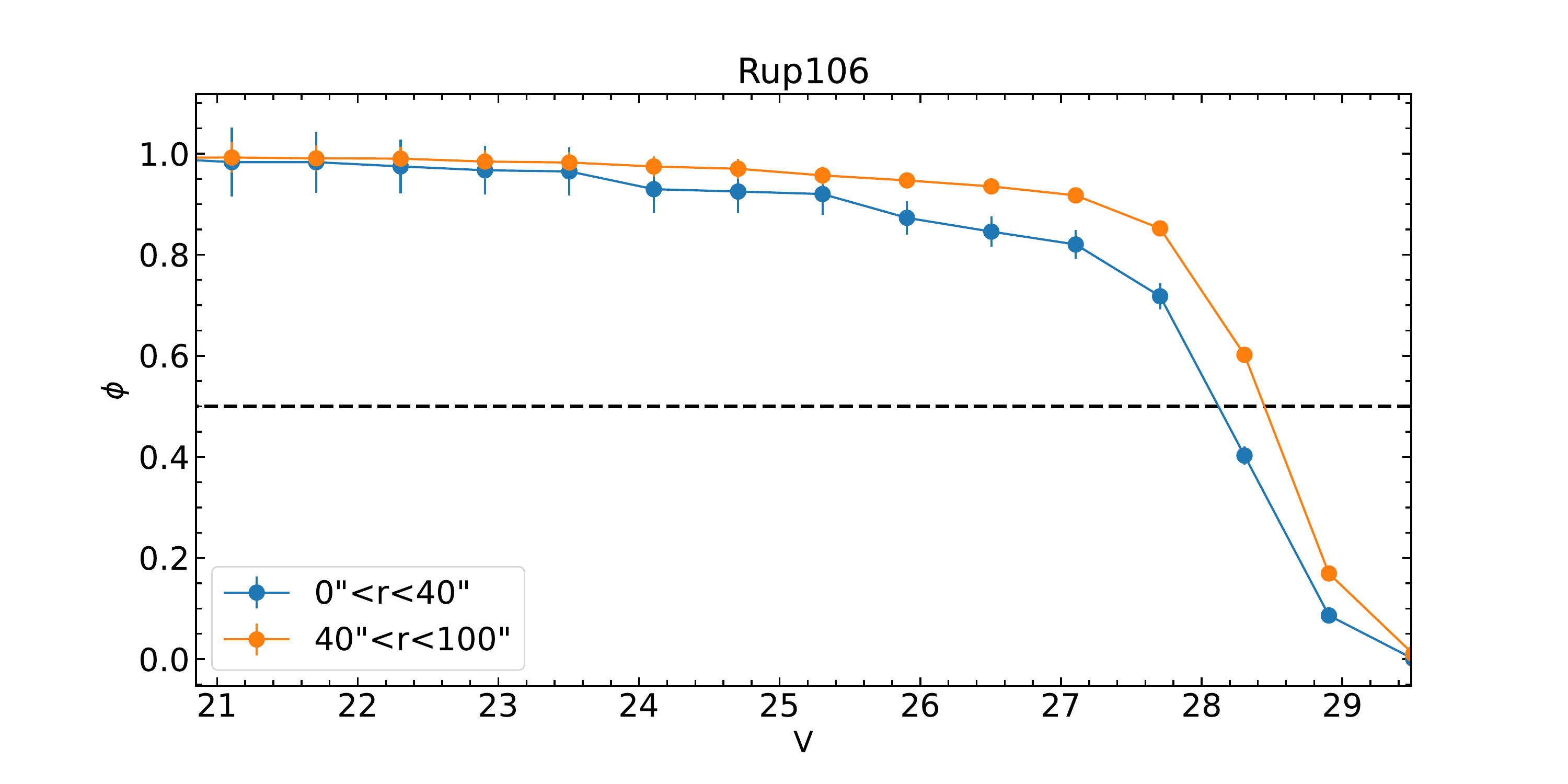}}
  \resizebox{0.85\hsize}{!}{\includegraphics{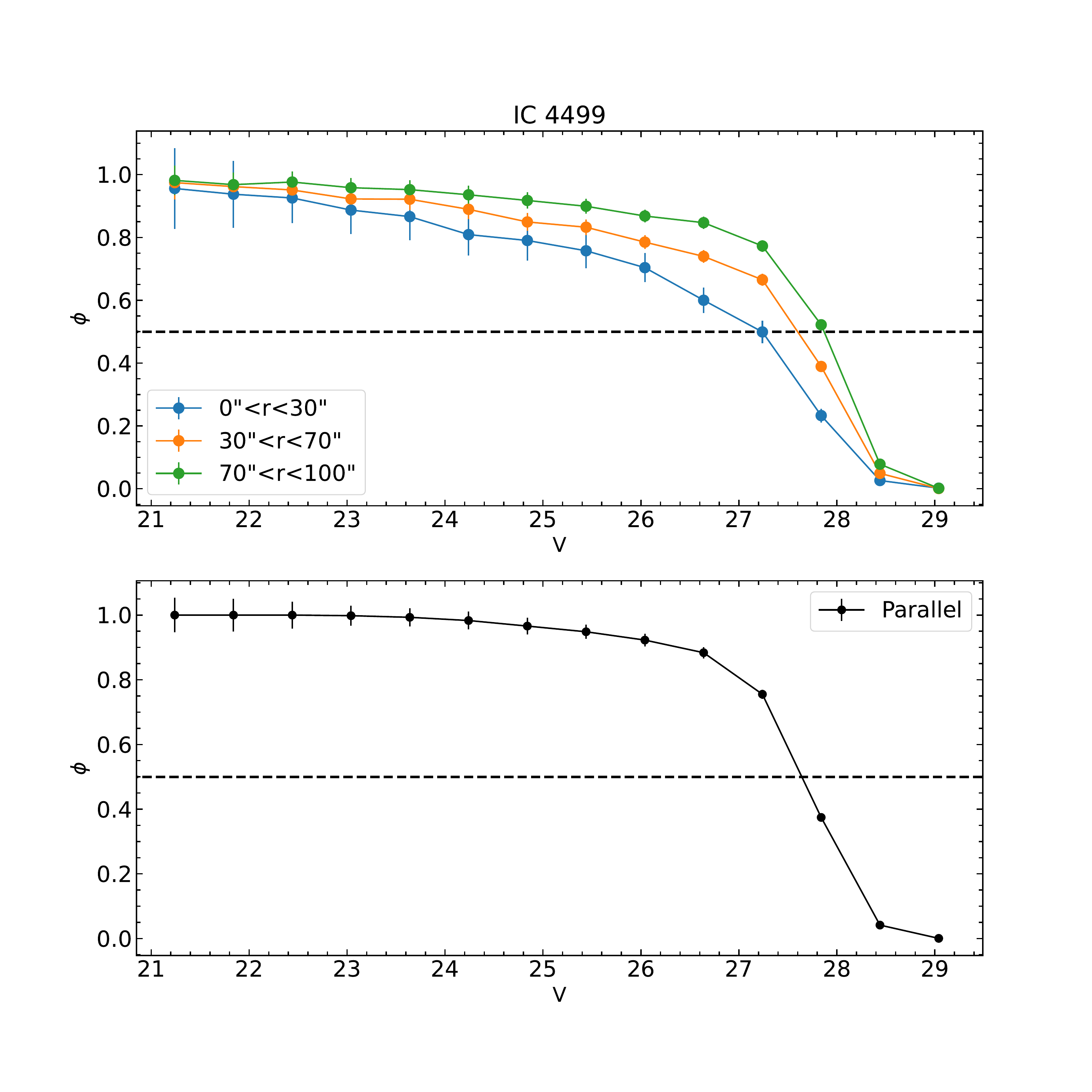}}
 \caption{Completeness. Photometric completeness ($\phi$) as a function of V for the ACS data of Rup~106 (upper panel) and for the ACS and WFC3 parallel observations of IC~4499 (central and lower panels).}
  \label{fig_comp}
\end{figure}


In Fig.~\ref{fig_comp} we show the photometric completeness of the data ($\phi$) calculated as the ratio between the number of artificial stars injected into the images and the ones recovered, as a function of the V magnitude and at different radial annuli from the clusters' centres. 
The radial ranges covered by the annuli are reported in arcseconds on the figure.  
The errors of the completeness value of each star ($\sigma_{\phi}$) were computed by propagation of the Poissonian errors. 
A value of $\phi=1$ indicate 100\% of recovery rate. We indicate with an horizontal dashed line the lower limit of 50\% (i.e. $\phi=0.5$) recovery rate below which the number of lost stars is too large to
safely correct the observed star counts (and derive the MF). 

The quality of the HST data used in this work 
is such that we detect in both clusters the MS stars in a
range of magnitude $24\lesssim$V$\lesssim21$ with an 100\% level of completeness while the completeness is always $>$80\% in the range  $27\lesssim$V$\lesssim24$. 

We use the accurate study of photometric completeness done for the HST data to assess the quality of the catalogues obtained from ground based photometry. As shown in Fig.~\ref{fig_comp}, the HST data are 100\% complete at magnitudes V$\lesssim21$. Hence, we can use all the stars sampled by the HST data in this regime of magnitudes as a reference catalogue and estimate how many stars with magnitude V$\lesssim21$ we recover with the ground based photometry. In particular we find that in the magnitude regimes $21\lesssim$V$\lesssim20$ and
 $20\lesssim$V$\lesssim19$  we respectively recover on the ground-based catalogues 80\% and 95\% of the stars sampled with the HST data with no significant difference with respect to the radial position. This is due to the fact that the completeness of the data is driven by the sensitivity toward faint magnitudes rather a sever stellar crowding, which is very low. This demonstrate that we can safely perform statistical studies based on star counts using both the HST and ground based catalogues.

\section{Ground based data-set: the cluster's structural parameters}
\label{sec_param}

We can take advantage of the accurate sampling of the stellar populations in the clusters
to determine the radial density profiles form the star counts. This can be done by following a standard
procedure already adopted by~\citet[][and reference therein]{lanz19}. In short, we
first used the ground based catalogues to estimate the clusters' centres calculated as the
average position projected on the sky of the sampled stars from V=15 down to V=20,20.5,21,21.5,22,22.5,23. Moreover, we only use those stars within a half-mass radius from the cluster's centre. At this stage, we adopted the values of the half-mass radius reported in~\citep[][2010 edition]{Harris2010}. The use of progressively fainter limiting magnitudes allows us to increase the number of sampled stars while mitigating possible effects related to the decreasing photometric completeness. The adopted centres are the average of the values obtained in the different magnitude bins and the dispersion represents the corresponding uncertainty. The coordinates of the centres are
$\alpha_{J2000}=12^{\rm h}\,38^{\rm m}\,40^{\rm s}.56, \delta_{J2000}=-51^\circ\,09^\prime\,01^\prime\,58$ for Rup~106 and $\alpha_{J2000}=15^{\rm h}\,00^{\rm m}\,16^{\rm s}.67, \delta_{J2000}=-82^\circ\,12^\prime\,49^\prime\,47$ for IC~4499. The overall uncertainties are $\sim2\arcsec$ and $\sim1\arcsec$ in $\alpha$ and $\delta$, respectively. 


\begin{figure*}[ht!]
  \scalebox{0.5}{\includegraphics{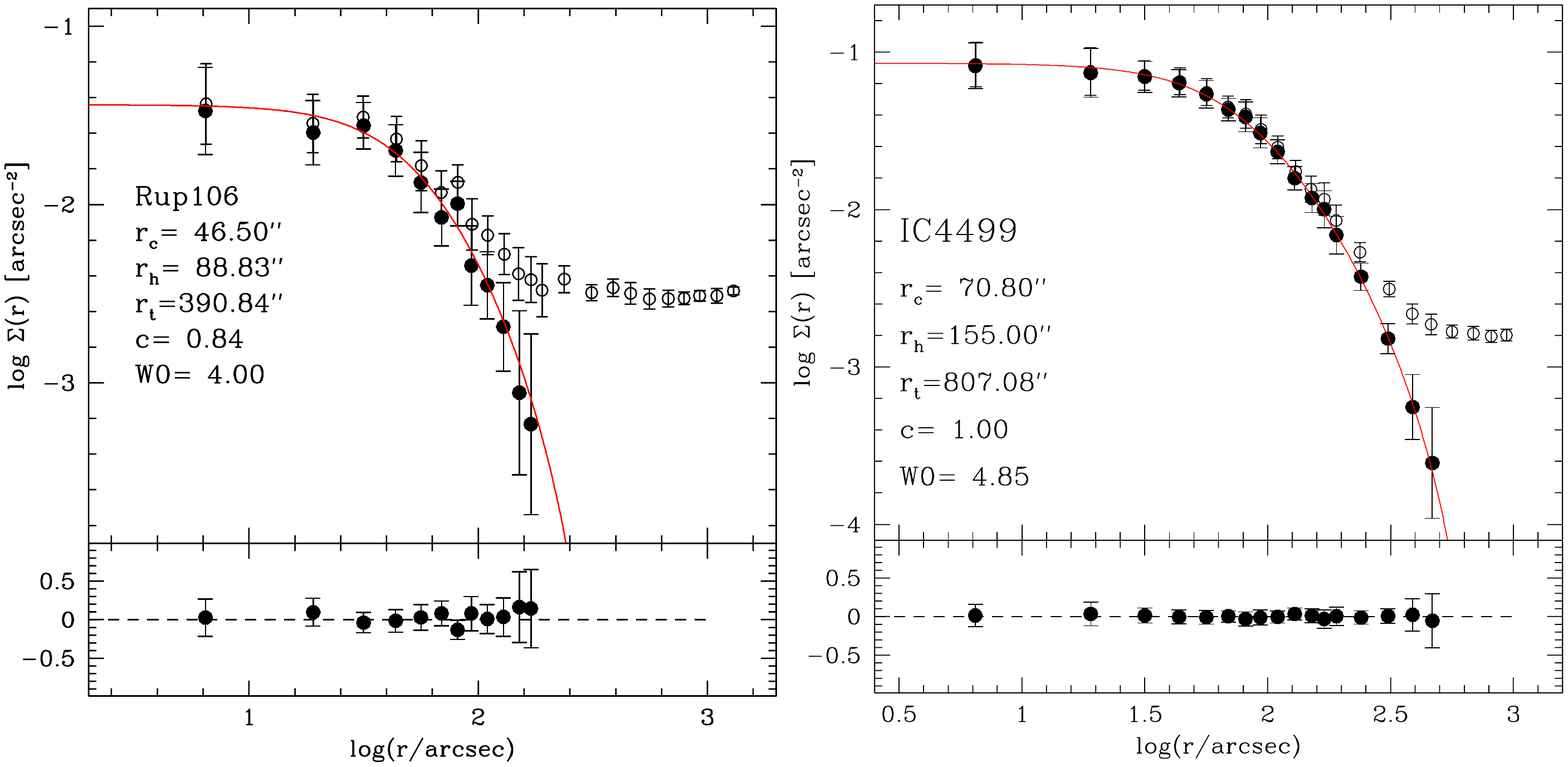}}
  \caption{Density profiles. The solid circles show the radial density profile of Rup~106 (left panel) and IC~4499 (right panel) corrected for the contamination from the back/foreground stars. The open circles indicate the back/foreground uncorrected counts. The density of field stars was estimated using the most external bins. The best fit King model is shown as a solid line on the figures while the small panel at the bottom of the figures shows the residual of the fit at each radial bin. The parameters of the best-fit King model are also indicated in the figure.}
  \label{fig_prof}
\end{figure*}


As a second step, we used the ground based catalogues to derive the radial density profiles of the two clusters from the star counts. We build the profile by calculating the density of the stars in several annuli, all centred on the clusters' centre. Specifically, we divide each annulus in four quadrants of identical size. The stellar density of the annulus is then the mean of the stellar density in the four individual quadrant while the standard deviation is taken as the associated error. We used all the stars with magnitude V<21. We show as open circles in Fig.~\ref{fig_prof} the observed density profiles of the clusters. The solid circles in the plots show the stellar density once the stellar back/foreground contamination has been removed. The latter has been estimated by using the 11 and 5 most external bins in the density plot for Rup~106 and IC4499, respectively.  

The background corrected profiles were fit with single-mass, spherical and isotropic King models \citep{King1966}, which constitute a single-parameter family, where the shape of the density distribution is uniquely determined by the value of the concentration parameter ($c$) or, equivalently, the value of $W_0$ that is linked with the dimensionless central potential. For each cluster, we compared the observed profile with a grid of King models with $c$ varying between 0.8 and 2.5 in steps of 0.05 \citep{Miocchi2006}. From the analysis of the residuals between the observed and the theoretical profiles, we then determined the best-fit solution as the model providing the minimum $\chi^2$ value (see \citealp{lanz19} for more details). The best-fit models are shown as solid red lines in Fig. \ref{fig_prof}, where we also labelled the value of the main structural parameters. In both cases, we find very small values of the concentration parameter ($c=0.84$ and 1.02 for Rup 106 and IC 4499, respectively) and large core radii ($r_c=46.5\arcsec$ or 4.10pc and $70.9\arcsec$ or 6.46pc, respectively), in agreement with the expectations for such low-density systems.
Moreover, Rup~106 has a tidal radius r$_t=390\farcs84$ (34.48pc) and for IC~4499 we find that r$_t=807\farcs08$ (73.56pc). We estimate the size of the radii in pc by assuming the distance of 21.20Kpc and 18.8Kpc for Rup~106 and IC~4499, respectively~\citep[][2010 edition]{Harris2010}.   

We notice that the structural parameters that we derive in this work for the two clusters slightly differ from 
the ones reported by~\citet[][2010 edition]{Harris2010} and the recent study done by~\citet[][]{Baumgardt18}. Arguably, the difference
is due to the data-sets used in our work and by~\citet[][]{Baumgardt18} to estimate the parameters. While our data allow us to sample the stellar population of the clusters well beyond the r$_t$, ~\citet[][]{Baumgardt18} used HST archive data to sample mostly the central regions complemented with surface brightness profiles to sample the external areas. 
We stress here that determining the radial density profiles from the star counts using properly resolved stars allow us to exclude any possible bias induced by the presence of a few bright star, that would significantly alter the location of the surface brightness maximum and possibly the precise shape of profile itself~\citep[see also][]{lanz19}. Moreover, as proven by ~\citet{Miocchi2013} the availability of wide fields catalogues able to sample the cluster's stellar population for its total extension is essential to have a solid control of the level of background contamination.

\section{Combined data-set: the radial distribution of the Blue Straggler Population}\label{sec_bss}

As extensively shown in the literature, the population of BSS and their radial distribution can be 
successfully used to investigate the dynamical age of a GGC. In GCs, BSSs are on average more massive~\citep[$\sim1.2$M$_{\odot}$; see e.g.][]{Fiorentino2014,raso2020} 
than the cluster's stars whose average mass is of the order of $0.3-0.4$M$_{\odot}$. 
As a consequence, BSSs sink to the centre because of dynamical friction. Indeed, when compared to the radial distribution of standard stellar populations in a GC (e.g. RGB), the BSSs are typically more centrally concentrated, suggesting that dynamical friction efficiently segregates BSSs toward the central potential well of the cluster~\citep[see e.g.][]{Ferraro2012}. 

\citet{f18} published a comparative study of the radial behaviour of the BSS population in
48 GCs adopting the parameter A$^+$. Initially introduced by~\citet{al16}, the A$^+$ of a GC 
is defined as the  area  enclosed  between  the  cumulative  radial  distribution  of the population of  BSS  of a given GC and that of a  reference population (e.g. RGB or MSTO stars). This dimensionless parameter turns out to be a solid empirical indicator of the degree of segregation of the population of BSS with A$^+$ increasing as a function of the degree of sedimentation of the
BSS into the cluster centre. The parameter A$^+$ tightly correlates with the ratio of cluster age to the current central relaxation time, making A$^+$ an efficient clock to empirically measure the dynamical age of GCs~\citep[see also][]{la16}.  


We aim here at using the BSS and the A$^+$ as an empirical clock to assess the dynamical age of Rup~106 and IC~4499. As a first step, we select the population of bona-fide BSSs in the CMDs of the clusters. 
The BSSs live along the MS in a position brighter and bluer with respect to the MSTO. Indeed, the BSS sequence of the two clusters are clearly visible in the CMDs shown in Fig.~\ref{fig_cmd}.


\begin{figure*}[ht!]
 \scalebox{0.3}{\includegraphics{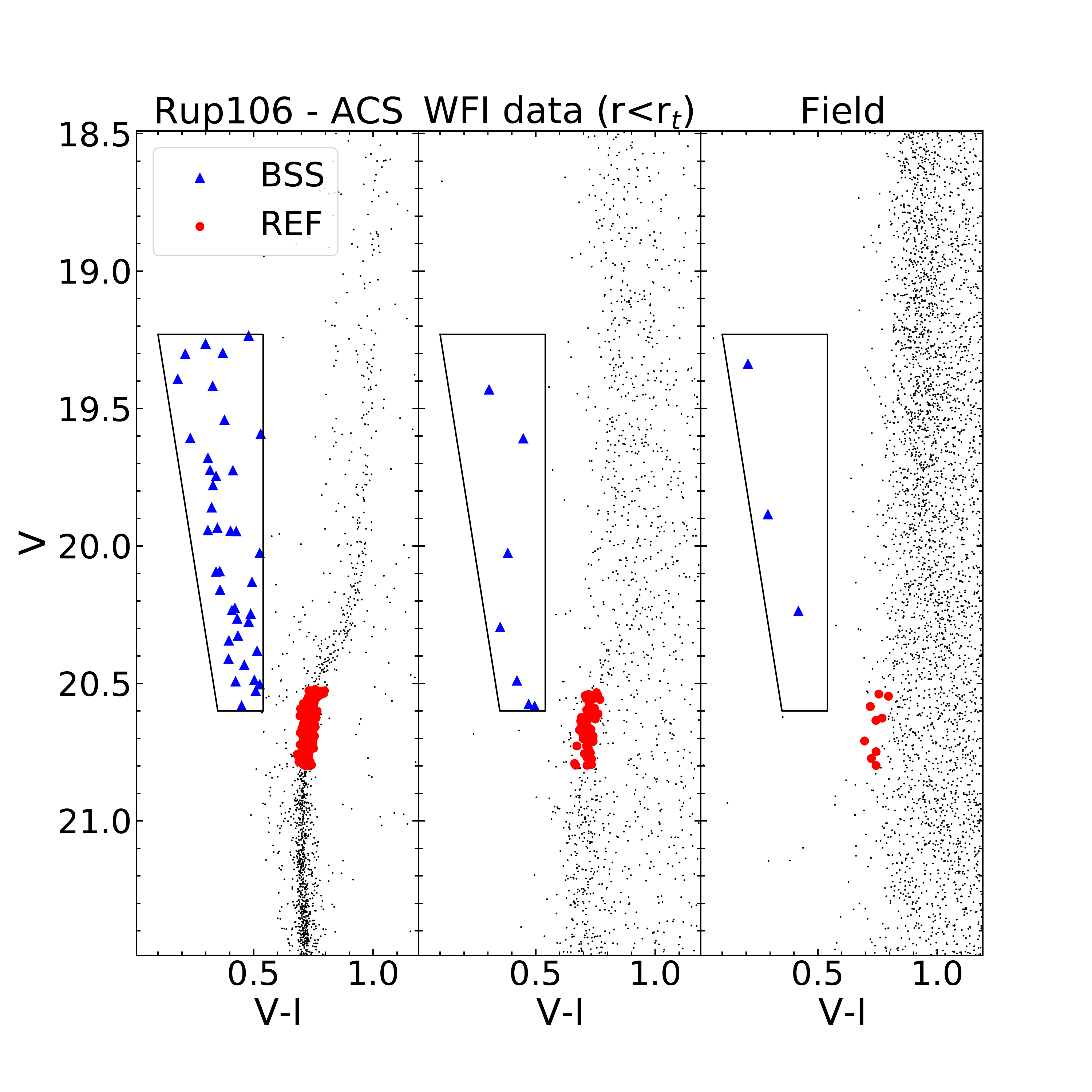}}
 \scalebox{0.3}{\includegraphics{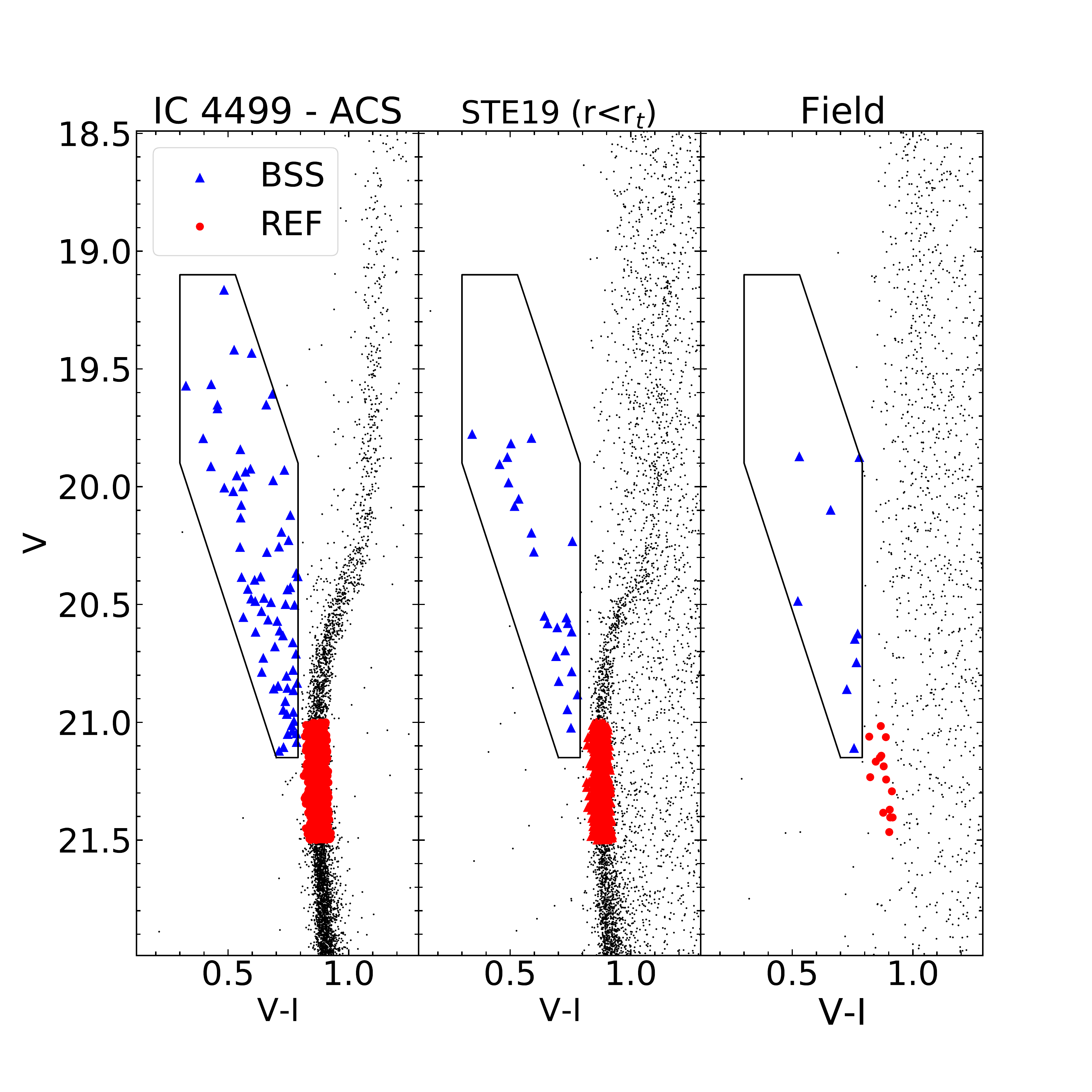}}
  \caption{Population' selection. Selection of the population of BSSs on the ACS and ground based data-sets of Rup~106 and IC~4499. The solid box empirically defines the BSS area. The selection boxes are homogeneously used to select BSSs  (blue solid triangles) on all the catalogues. The stars along the MS turn-off (red solid circle) are used as reference population.}
  \label{fig_sel}
\end{figure*}


Following the definition of the population of BSSs given above, we empirically define a BSS selection box for Rup~106 and IC~4499 as shown
in Fig.~\ref{fig_sel}. The populations of BSSs in the two clusters are shown as blue solid triangles in the figures while we adopted the stars at the MSTO (REF; red solid circles in the figure) as reference population sampling the distribution of the light in the clusters. 
The selection is homogeneously applied across the entire available data-sets 
thanks to the fact that, as described in Sec.~\ref{sec_red}, for both GCs studied in this paper, the V and I magnitudes of all catalogues have been reported to the same photometric system. 
We selected a total of 56 BSS and 745 REF stars inside the r$_t$ of Rup~106, while 98 and 1896 are found in IC~4499.
In addition, given the radial
extent of the ground-based data-sets, 
we are able to provide a solid estimate of the contamination by fore/background of BSS and REF stars, which turn out to be negligible for both clusters - less than one field star inside the r$_h$ is expected among the population of BSS and REF stars for both clusters. We would like to remark here that the MSTO of the clusters are located at V$\sim21$mag, i.e. they are too faint to allow us to use the Gaia early Data Release 3 catalogue to asses individual cluster membership based on robust astrometric information for the stars populating the BSS sequence.


\begin{figure*}[ht!]
 \scalebox{0.3}{\includegraphics{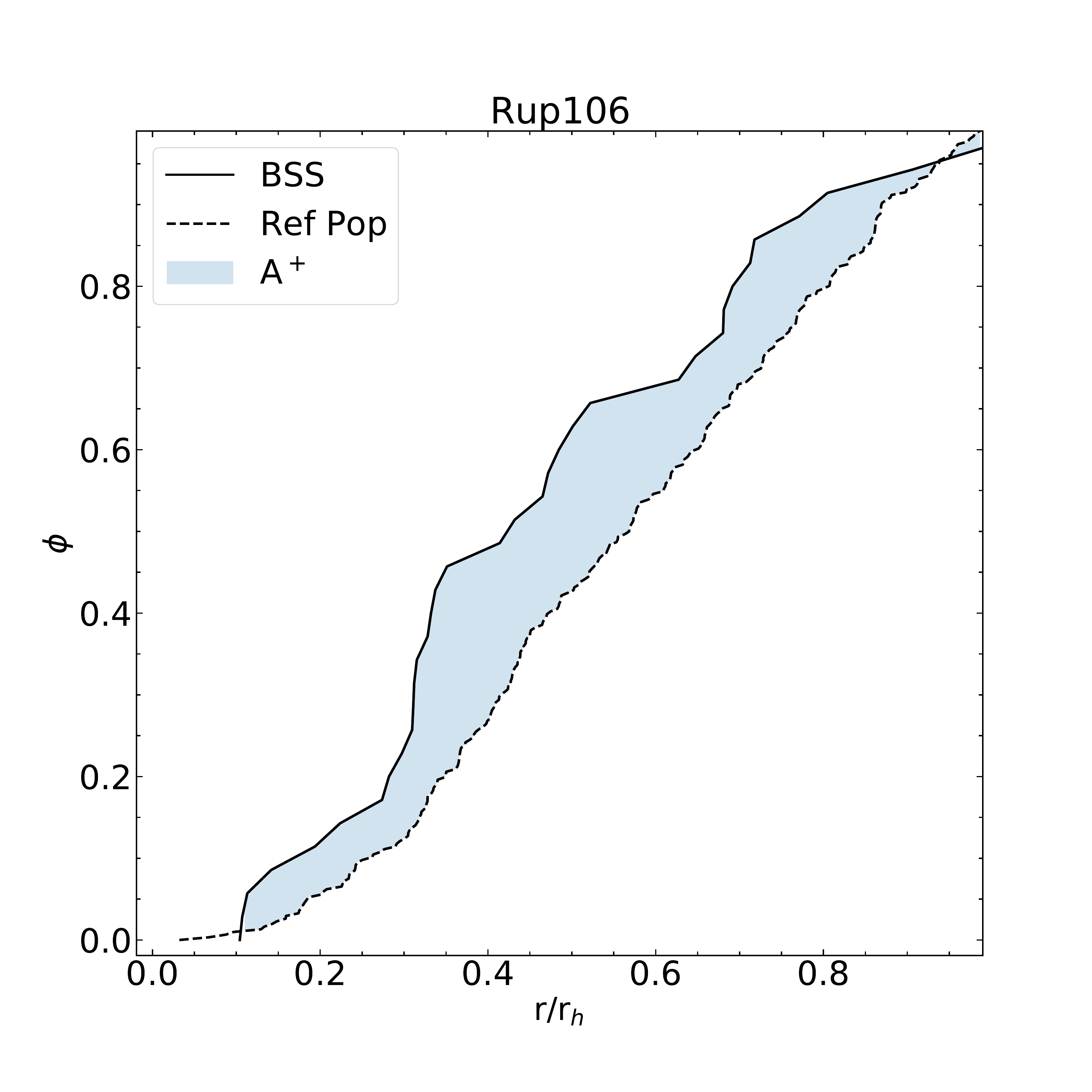}}
 \scalebox{0.3}{\includegraphics{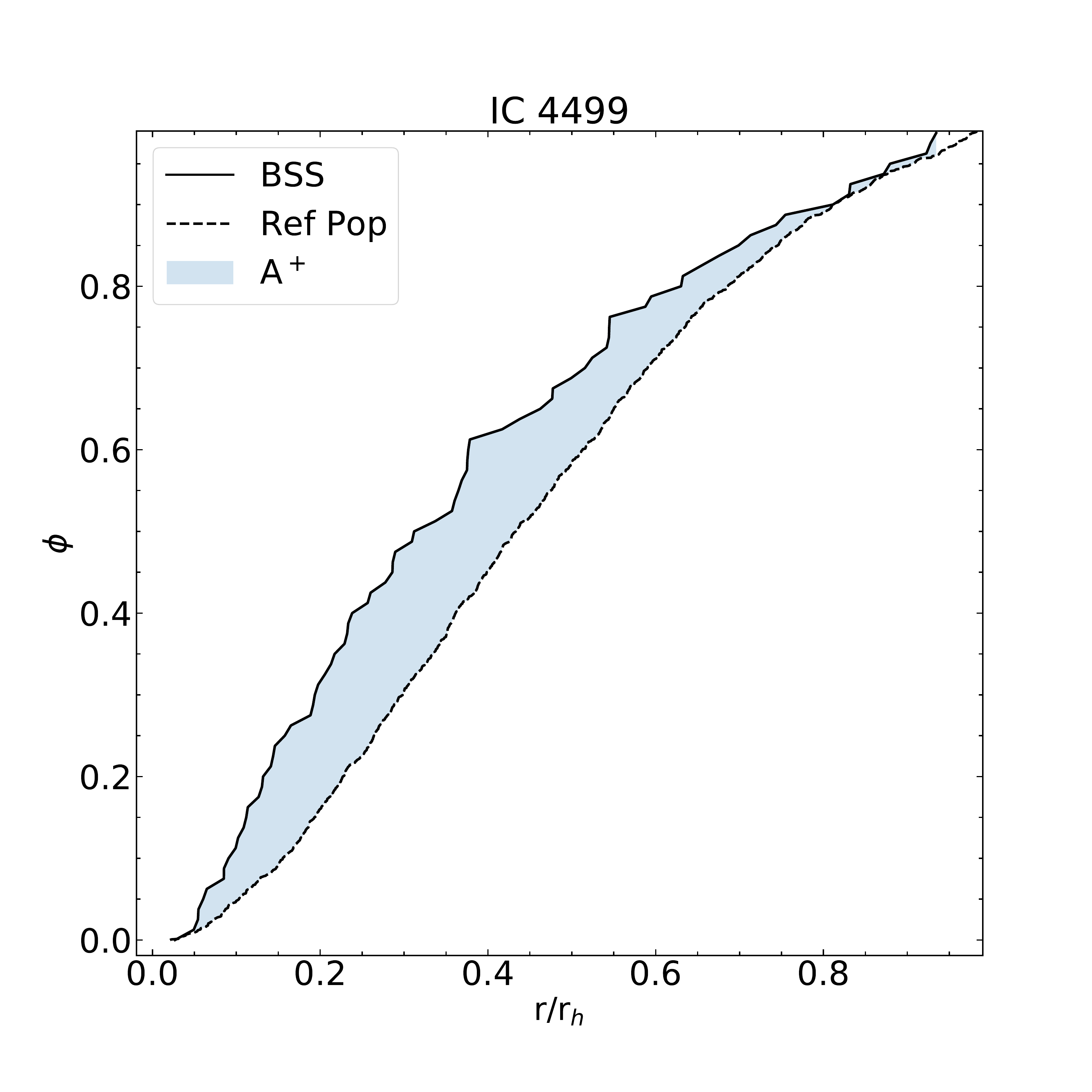}}
  \caption{Estiamtion of A$^+$. Cumulative radial distributions of the stars in Rup~106 and IC~4499 (left and right panels, respectively) belonging to the population BSS and the reference one (REF), the latter taken as representative of the distribution of the light of the clusters. The radial distributions are normalised to the clusters' r$_h$.}
  \label{fig_rad}
\end{figure*}


We show in Fig.~~\ref{fig_rad} the cumulative radial distribution of the population of BSS and REF stars in the cluster Rup~106 and IC~4499 (left and right hand panels, respectively). The cumulative distributions are derived with respect to the centre of each clusters obtained in Sec.~\ref{sec_param} and the distances are normalised to the clusters' r$_h$.  
We computed the Kolmogorov-Smirnov statistic on the samples of BSS and REF stars in both clusters to test whether the 2 samples are drawn from the same parent distribution. 
The KS test gives a p-value in both cases well below  $1\%$. This fact implies that we can safely reject the hypothesis that the distributions of the BSS and the REF samples comes from the 
same parent population.

In order to further quantify the difference in the populations' distributions, we calculate A$^+$ following the recipe by \citet[][]{f18}. In particular, the choice of estimating the A$^+$ when the radial distribution are normalised to the clusters' r$_h$, makes it possible to perform a direct comparison of the A$^+$ of Rup~106 and IC~4499 with the one of the clusters studied in  \citet[][]{f18}.
We find A$^+$ to be $0.111\pm0.027$ and $0.079\pm0.011$ for Rup~106 and IC~4499, respectively.
Errors on A$^+$ have been obtained by using jackknife bootstrapping technique
(\citealt{lupton93}; see also \citealt{dalessandro19}). 
The A$^+$ values derived in this work for Rup~106 and IC~4499 can then be directly compared to the large compilation of clusters studied in \citet[][]{f18} to constrain the dynamical age of the target clusters.


\begin{figure}
  \scalebox{0.3}{\includegraphics{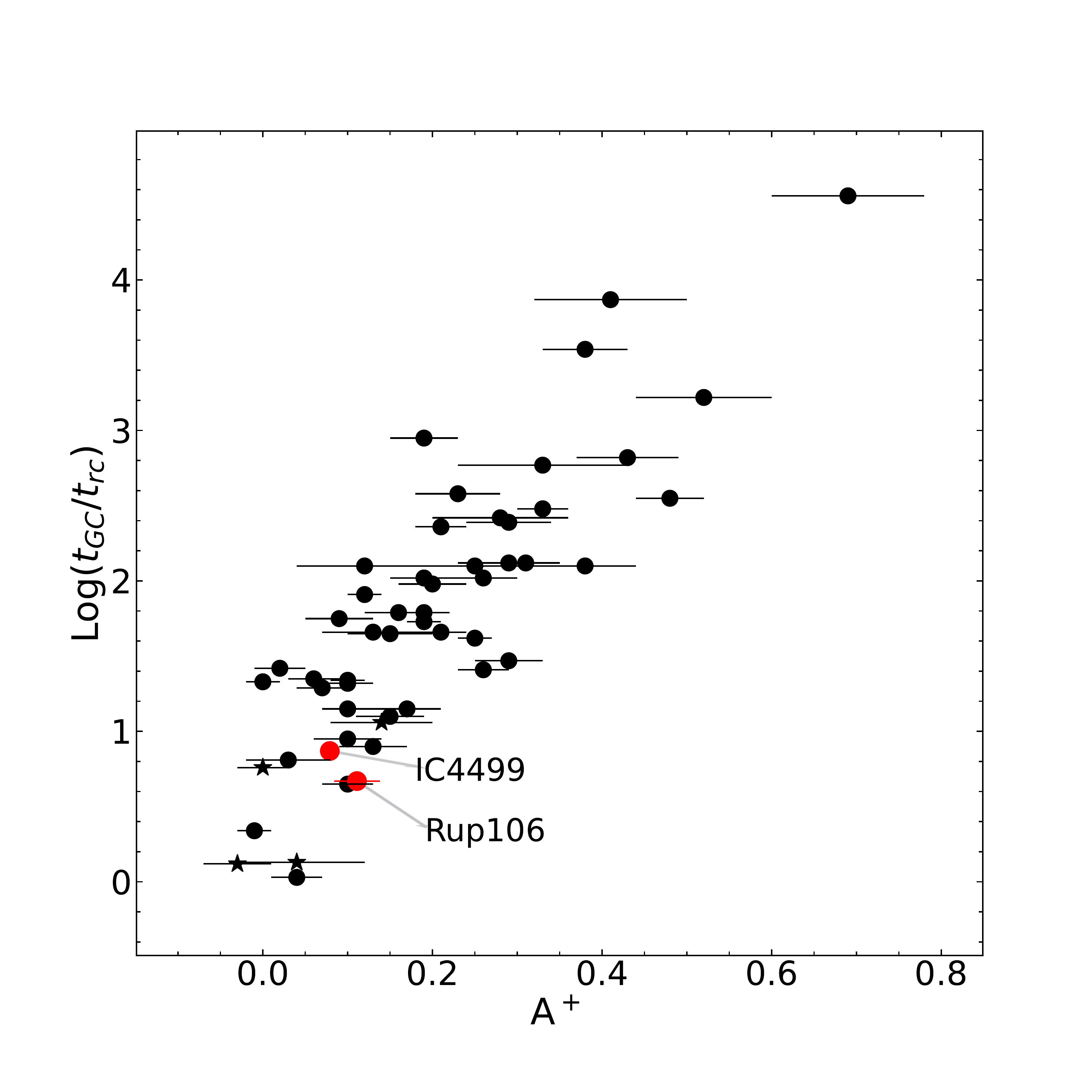}}
  \caption{Dynamical clock. The value of A$^+$ calculated for Rup~106 and IC~4499 (red solid circles), the GC studied in~\citealt{f18} (black filled circles) and the Halo GCs AM1, Eridanus, Pal 3 and Pal 4 (solid star-like points) studied in~\citet[][]{beccari12} as a function of the logarithm of the cluster core relaxation time (t$_{rc}$), that have occurred since the epoch of cluster formation t$_{GC}=12$Gyr.}
  \label{fig_aplus}
\end{figure}


We compare in Fig.~\ref{fig_aplus} the values of A$^+$ for the 48 GCs published in~\citet[][]{f18} (black solid circles) and Rup~106 and IC~4499 (red solid circles) against the number of core relaxation times ($t_{rc}$) that have occurred since the epoch of cluster formation ($t_{GC}=12$Gyr): N$_{relax}$ = t$_{GC}$/t$_{rc}$.

We add on the same plot the  A$^+$ obtained using the BSS population of the Halo GCs AM1, Eridanus, Pal 3 and Pal 4 (solid star-like points) published in~\citet{beccari12}. 

As already explained in~\citet[][]{f18}, we observe a clear correlation between the value of A$^+$ and N$_{relax}$ with the value of A$^+$ of a GC increasing as the cluster evolves dynamically. This is clearly demonstrated by the fact that the cluster at the top right of this plot, i.e. the one with the biggest value of A$^+$, is M~30 which has shown empirical signatures of having experienced the collapse of the core  ~\citep[see][]{fe09}. On the contrary, the cluster with the smallest value of A$^+$ at the bottom left of Fig.~\ref{fig_aplus} is OmegaCen, which has been shown to not be fully relaxed even in in the central regions~\citep[][]{Ferraro2006}.

Clearly the  values of A$^+$ of Rup~106 and IC~4499 nicely follow the observed trend. Interestingly, the four Halo GCs studied in~\citet{beccari12} and shown on this plot as solid star-like points, share similar properties with respect to Rup~106 and IC~4499, being all younger (10 to 11Gyr) and 
slightly less massive than the bulk of  the GC in the Galaxy ~\citep[see][for a detailed discussion]{beccari12}. 
The plot in Figure~\ref{fig_aplus} demonstrates that the dynamical history of this family 
of GCs as imprinted in the radial distribution of the BSS nicely resemble the one of whole family of Galactic GCs. While we will discuss later the
possible implication of this result with respect to the origin of these GCs, it is worth here to mention that, given the large distance of these Halo GCs (>50 kpc from the Galactic centre) none of the current observing facility would allow us to sample the MS mass function to study the degree of mass segregation in these clusters.
The  A$^+$ is hence the only empirical method which allows us to investigate the dynamical age of such remote GCs.

More specifically, the A$^+$ values derived for Rup~106 and IC~4499 result to be in very good agreement 
with those of dynamically young clusters. 

\section{The deep sample: The radial distribution of the Mass Function}
\label{sec_mf}


To further investigate the mass segregation phenomenon in
Rup~106 and IC~4499, we analysed the radial variation of the MF of the MS
stars. The study of the shape of the MF in different regions of the clusters gives 
information about the effect of cluster internal
dynamics on stars in a wide range of masses, including the
faint-end of the MS where most of the cluster mass lies. In
relaxed stellar systems, the slope of the MFs is expected to vary
as a function of the distance from the cluster centre, with indexes decreasing as the 
distance increases, because of the differential effect of mass segregation.


As shown in Fig.~\ref{fig_maps}, 
the ACS images alone allow us to fully sample the core region in both the clusters.
Ideally, in order to fully appreciate the change in the
slope of $\alpha$, it is important to derive the MF for a as large as possible fraction of the cluster extension and 
always beyond the half-mass radius.
In the case of Rup~106, the ACS images sample the cluster out to the half-mass radius while the latter 
falls outside of the ACS field of view in the case of IC~4499 (see Figure~\ref{fig_maps}).  
We hence used photometry obtained from a set of HST WFC3 parallel observations (Section~\ref{sect_ObsData}) in 
order to sample the MS stars in IC~4499 beyond the half-mass radius.



\begin{figure*}[ht!]
 \scalebox{0.3}{\includegraphics{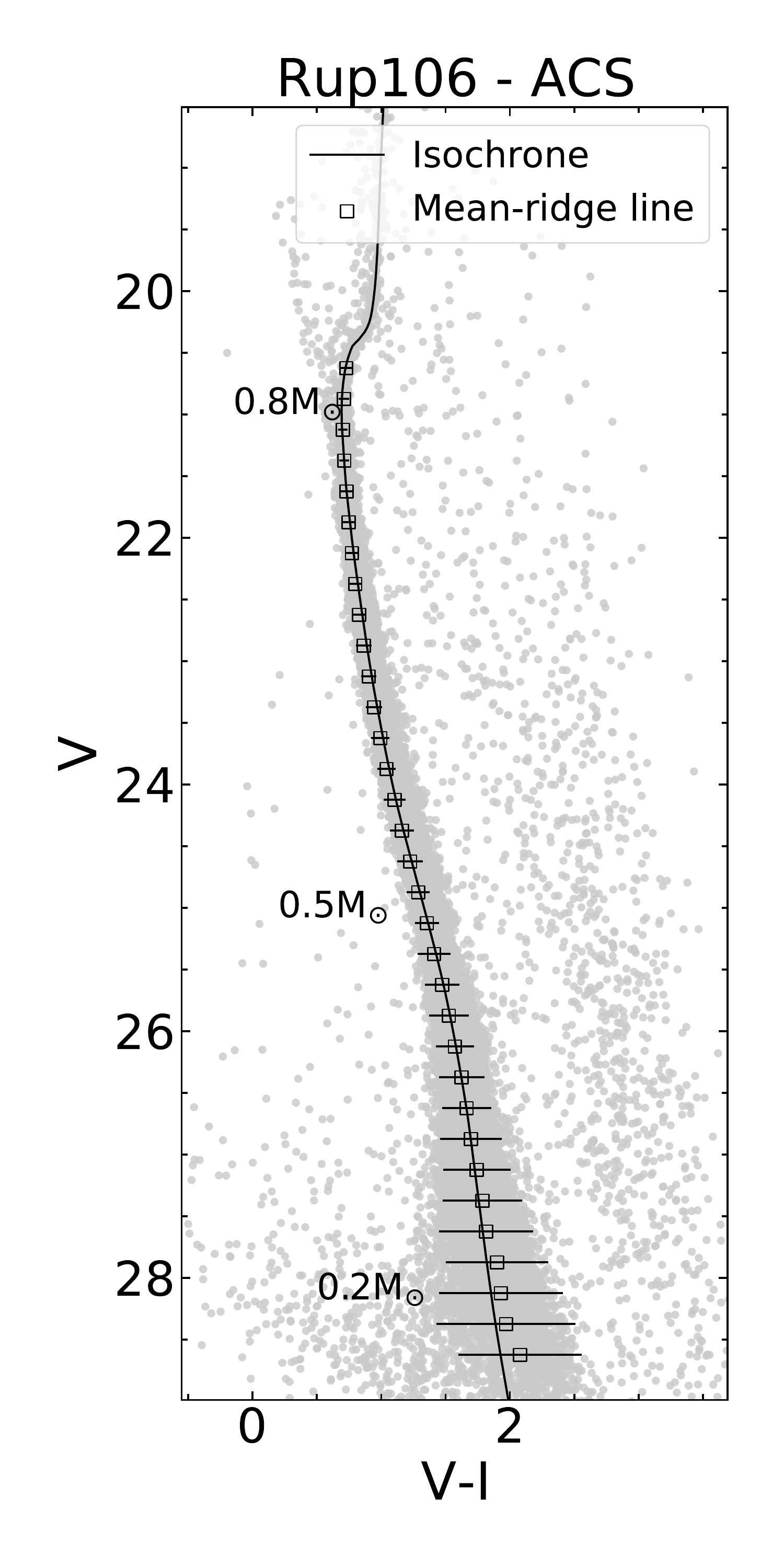}}
 \scalebox{0.3}{\includegraphics{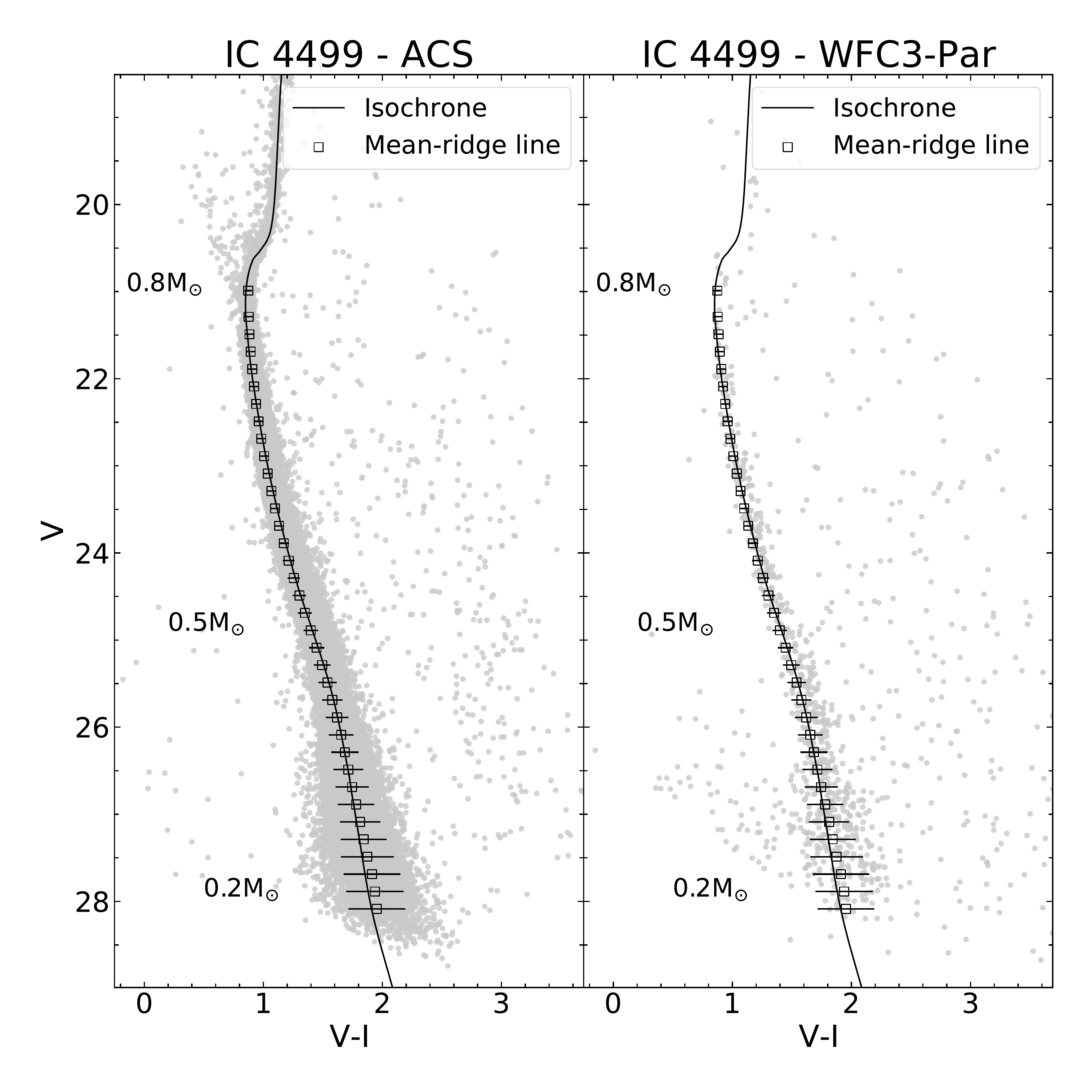}}
  \caption{CMD of the HST data for Rup~106 and IC~4499 (left and right bi-panels, respectively). The ACS data used to sample the core region are complemented with a catalogue obtained with ACS and WFC3 parallel observations for Rup~106 and IC~4499, respectively. The open squares indicates the clusters' mean ridge lines, while the solid curve shows the best-fit isochrones.}
  \label{fig_iso}
\end{figure*}

The HST data used in this
paper were used by~\citet{Dotter2011} to study the age and metallicity
of six halo GCs, including Rup~106 and IC~4499. The authors
performed a detailed study of the age of the clusters via isochrone fitting, following the recipe from~\citet{Dotter2010}. 
In our work we hence adopted the models from~\citet{Dotter2008} and the best-fit isochrone parameters reported in Table 3 of~\citet{Dotter2011}. 
For completeness we report the parameters here in Tab.~\ref{tab_iso}.
We show in Fig.~\ref{fig_iso} the best-fit isochrones for both clusters (black lines) and
with the open squares we indicate
the mean-ridge line of the MS and the associated 1$\sigma$ uncertainty. 
Fig.~\ref{fig_iso} clearly demonstrates the exquisite fits of the MS obtained by using the mentioned isochrones and distance and reddening values. 
We also highlight in the figures the range of masses covered by the data. 
 As discussed in Sec.~\ref{sect_CompTest} the quality of the adopted images allow us to sample the MS stars with a photometric completeness better than 80\% down to V=27.5 mag. This correspond to a capability of sampling the MS
in a broad range of masses, down to the very low-mass regime of $0.25$M$_{\odot}$ (see Fig.\ref{fig_iso}).

\begin{table}
\centering
\caption{Best-fit parameters of the isochrone fitting from~\citet{Dotter2011}}
\label{tab_iso}
  \begin{tabular}{@{}lcclcl@{}}
  \hline
   Cluster   &   (m-M)$_\mathrm{V}$  & $E(B-V)$ & [Fe/H] & [$\alpha$/Fe] & Age (Gyr) \\
 \hline
IC~4499      & 17.07  &  0.18 & $-1.6$  &   0.2  &  $12.0\pm0.75$ \\
Rup~106 & 17.12  &  0.14 & $-1.5$  &   0.0  &  $11.5\pm0.5 $ \\
\hline
\end{tabular}
\end{table}


\subsection{The clusters' Mass Functions}

The MF is calculated using the stars along the clusters' MS. We selected all the stars
laying at a distance from the MS mean-ridge lines in the V-I colour $<2\sigma$ of the uncertainty of the line itself.
This approach allows us to exclude the majority of unresolved binaries, which are well known to 
populate a $secondary$ sequence parallel to the MS at a maximum distance of 0.75 magnitudes in the V-I colour~\citep[see e.g.][]{Sollima2008}.
We note that a fraction of binaries might indeed still be included in the selection. 
Still, those are binaries
with a very low mass ratio and whose luminosity is entirely dominated by the primary which is more massive. Hence there is no significant impact of unresolved binaries on the derivation of MFs. 

Using the detailed information given in Fig.~\ref{fig_comp}, we assigned to each star of a given V mag and position on the sky, a completeness factor $\phi_{\star}$ by interpolating the appropriate completeness curves.
Then, we use the mass-to-light conversion available on the adopted best-fit isochrones from~\citet{Dotter2008} to 
convert the V magnitudes of the stars previously selected along the MS, into stellar mass. 


\begin{figure*}[ht!]
 \scalebox{0.3}{\includegraphics{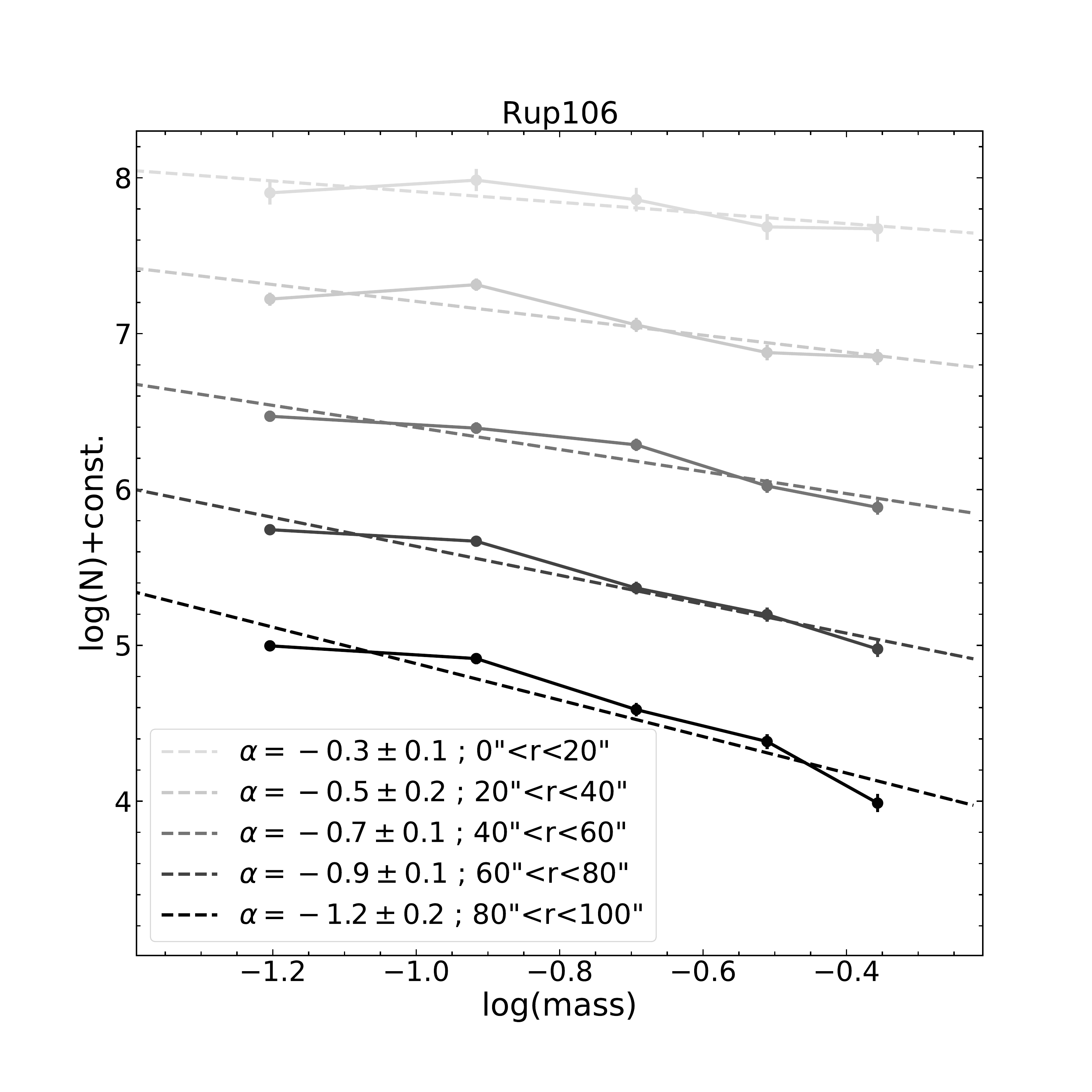}}
\scalebox{0.3}{\includegraphics{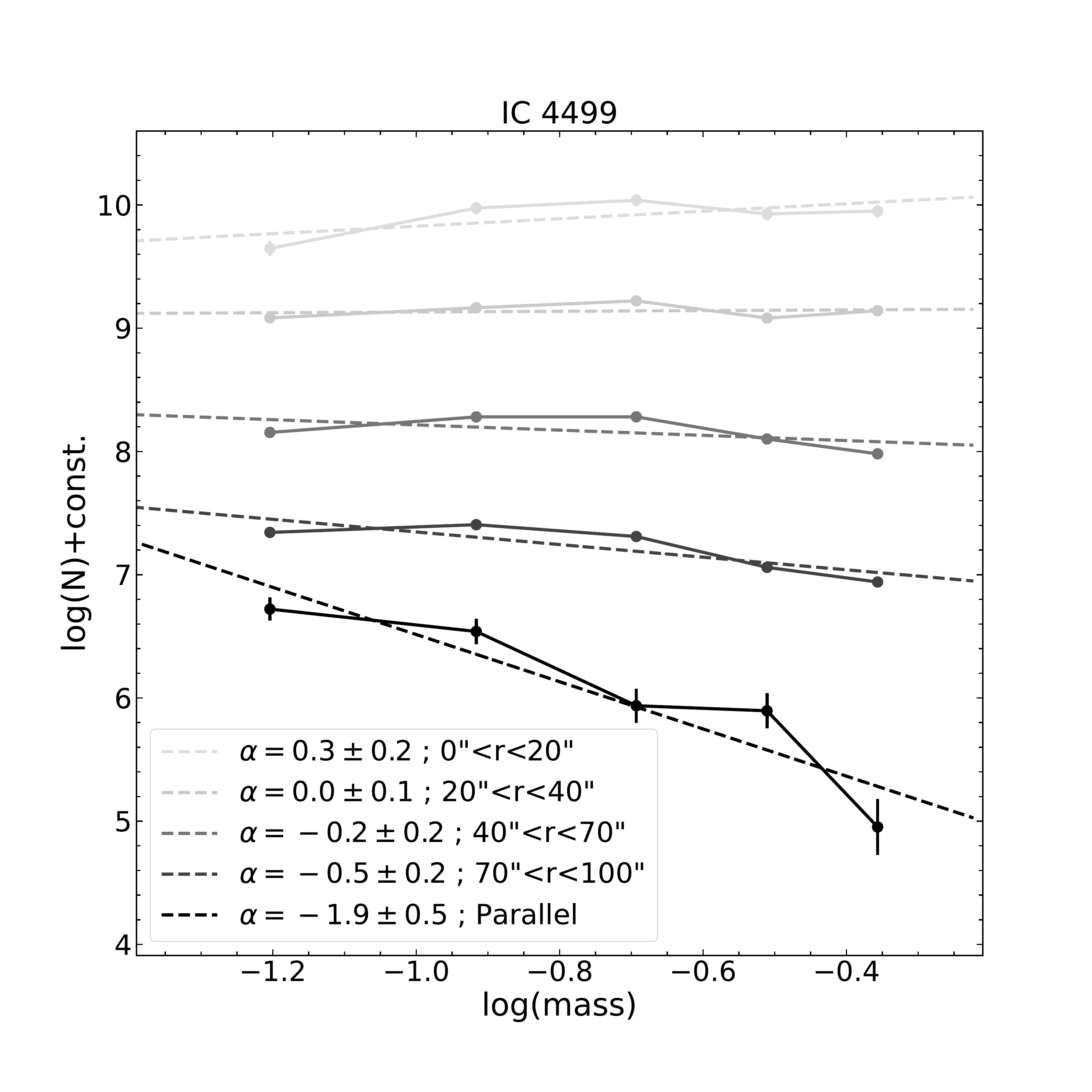}}
  \caption{Cluster's completeness corrected Mass Functions (MF) calculated at different radial annuli from the clusters' centres. The dashed line indicate the fit to the MF. The slopes of the fits indicate the slope of the MF and are also indicated in the plots. We remind here that in our notation, the Salpeter IMF would have a slope $\alpha=-2.35$, and that a positive index implies that the number of stars decreases with decreasing mass.}
  \label{fig_MF}
\end{figure*}


We binned the stars in the stellar masses range $0.25\leq$M$_{\odot}\leq0.75$ adopting a step of 0.1M$_{\odot}$.
The total number of stars in a given bin of mass is calculated as N$_{corr}=\sum_i^{N_{obs}}\phi_{\star_i}^{-1}$ while
the associated error is $\sigma_{N_{corr}}=\sqrt{\sum_i^{N_{obs}}(\sigma_{\phi_{\star_i}}/\phi_{\star_i})^2}$. This way we are able to build a completeness corrected
MF taking into account the incompleteness effects. In order to estimate the level of contamination on each bin of mass from background/foreground stars we downloaded from the Besan\c{c}on Galaxy Model~\citep[][]{Robin2003} a catalogue of
simulated stars in the V and I Johnson-Cousin magnitudes, sampling the Galaxy stellar population in a 1 deg$^2$ area around the two clusters. We selected the synthetic stars along the clusters' MS as done with the real stars
and converted their V magnitudes into masses using the same best-fit isochrones described above. Then, for each mass bin
in a given annulus where the MF was estimated, we removed from the N$_{corr}$ the number of contaminants in the
given mass bin scaled to the sampled area.

We show in Fig.~\ref{fig_MF} the final MFs for each cluster measured inside concentric rings sampling different distances.
The best fit to the MFs is shown as dashed line.
The slopes of the fits are used as the slope of the MF and are also reported in the plots. We remind here that in our notation, the Salpeter IMF would have a slope $\alpha=-2.35$, and that a positive index implies that the number of stars decreases with decreasing mass.
In the case of Rup~106 $\alpha$ ranges from -0.2 in the core of the
cluster to -1.2 in the most external region sampled with the ACS data, i.e. up to $100\arcsec$ or $\sim2$r$_c$.
We note here that the radial extension of the external annulus is such that it includes the r$_h=88\farcs83$ 
of the cluster. The radial behaviour of the MF in IC~4499 changes from $\alpha=0.3$ in the centre to -1.9 in the external region. 

Clearly in both clusters, the slope of the MF $\alpha$ significantly changes as a function of the distance from the clusters' centres.  Because of equipartition of energy, massive stars tend to have a smaller velocity dispersion than the average and, since they occupy smaller orbits, are found closer to the cluster centre generating mass segregation. The value of $\alpha$ becomes more negative while moving towards the external regions of the clusters as a consequence of the fact that the number of massive stars with respect of low mass stars is higher in the centre of the clusters. This fact confirms that a certain level of mass segregation is detected in both clusters. This result is qualitatively in agreement with what found by using the BSS radial distribution.

\section{Discussion And Conclusion}
\label{sect_Discussion}

In this work we have used a combination of high resolution HST data and photometric catalogues from ground-based wide-field telescopes to study the dynamical state and age of the globular clusters Rup~106 and IC~4499. In particular, using ground based catalogues combined with short exposures HST images of the core of the clusters, we derived the radial density profiles of the clusters from detailed star counts (see Fig.~\ref{fig_prof}). We show that the profiles are well reproduced by King models and we provide new estimates of the clusters' structural parameters.

In addition, we use the same catalogues to homogeneously select the population of BSS stars over the total extension of the clusters. We compared the radial distribution of the BSSs to that of a selection of stars along the MS turn-off , used as reference population (REF), i.e. a population sampling the normal distribution of the stars in the clusters. We used the parameter A$^+$, defined as the area between the cumulative distribution of the BSS and the REF populations (see Fig.~\ref{fig_rad}), to empirically constrain the dynamical age of Rup~106 and IC~4499. In Fig.~\ref{fig_aplus} we show the distribution of the A$^+$ of Rup~106 and IC~4499 together with the ones of several galactic GCs studied by~\citet{f18} as a function of the clusters' relaxation times at their core radius. This plot allow us to evaluate empirically the dynamical age of the two clusters studied in this paper. Fig.~\ref{fig_aplus} reveals that in both Rup~106 and IC~4499 two-body relaxation has only very recently started to efficiently sink the massive stars towards the central regions. Hence both clusters can be classified as dynamically young. 

The radial variation of the slopes of the clusters' MF $\alpha$ measured in the mass range $0.25\leq$M$_{\odot}\leq0.75$ also indicates that two-body
relaxation  has efficiently started to affect the distribution of the stars in the clusters. 
In particular, the value of $\alpha$ changes from $-0.1\pm0.1$ and $0.3\pm0.2$ in the central area of Rup~106 and IC~4499 respectively, to $-1.2\pm0.1$ and $-1.9\pm0.2$
in the external regions. As explained in the previous section, a progressively more negative (positive) $\alpha$ slope indicates the decrease (increase) of the number of massive stars with respect of low mass stars. Hence, the radial change of $\alpha$ indicates the presence of a certain degree of mass segregation in both clusters. It is worth to mention that in the case of Rup~106, we are able to sample the MF out to $\sim2\times r_c$ reaching the cluster's $r_h=88\farcs{83}$ while in the case of IC~4499 $\alpha$ is estimated out to $\sim5\times r_c$ or $\sim2\times r_h$.  

The nice agreement between the different mass segregation indicators used in this work for Rup~106 and IC~4499 (see also \citealt{bellazzini12,dalessandro2015} for the cases of NGC~2419 and NGC6101) provides further support to the use of the radial distribution of BSS as a powerful indicator of the dynamical evolution of stellar
systems (e.g. \citealt{Ferraro2012,f18}). In fact, the use of BSSs is quite convenient as they are significantly brighter
with respect to the MS stars, they are easy to identify with high level of completeness and with short exposure times
and their use to asses the degree of mass segregation (the radial distribution) is free from theoretical assumptions from
stellar evolution models. Hence the BSS radial distribution represents
the clearest indicator of mass segregation in dense stellar systems.
The plot in Fig.~\ref{fig_aplus} not only indicates (in agreement with the radial distribution of $\alpha$) the presence of mass segregation, but it also demonstrates that two-body relaxation has only very recently started to efficiently act in Rup~106 and IC~4499. 

Our study indicates that, from a strictly dynamical perspective,  the behaviour of both clusters is identical when
compared to those of the GCs in the Milky Way (MW; see Fig.~\ref{fig_aplus}). 
As mentioned in Sec.~\ref{sec_bss}, the A$^+$ of the Halo GCs AM1, Eridanus, Pal 3 and Pal 4 (solid star-like points in Fig.~\ref{fig_aplus}), indicates 
that all these clusters show a dynamical history (as imprinted in the BSS radial distribution) that well compares to the ones of any other Galactic GCs. 
Specifically, while all these clusters are slightly younger (10-11Gyr) in terms of the absolute age of the stellar population and with respect to the majority 
of GCs in the Galaxy, they are all also young from a dynamic perspective, as shown by their location on the plot of Fig.~\ref{fig_aplus}.

Interestingly,~\citet[][]{dalessandro18} discovered the presence of a mild 
spread in N and Na abundance among the the RGB of IC 4499 interpreted as the signature of the presence of multiple-stellar populations
in the cluster.  While it is well established that the 
stellar populations in GCs exhibit intrinsic star-to-star variations in their light element 
abundances~\citep[see e.g.][ and references]{carretta21}, it is still not clear why Rup~106 does not show any chemical signature that would indicate the presence of multiple-stellar populations~\citep[][]{fuj2021}. Currently, despite the efforts, no spectroscopic studies have
been able to prove the presence of multiple stellar populations in the other Halo GCs mentioned above
on a statistically solid base~\citep[see e.g.][]{koch2009,koch2010}.
Clearly, our study excludes that a peculiar dynamical history might have played a role in the luck of multiple stellar populations in Rup~106.  

Interestingly,~\citet[][]{Baumgardt18} indicates that IC~4499 is, among the mentioned group, the GCs with the highest initial total mass M=$10^{5.53}$M$_{\odot}$. Rup106 has a predicted initial total mass of $10^{5.12}$M$_{\odot}$ while AM1, Eridanus, Pal 3 and Pal 4 have a initial total masses lower 
than $10^{5}$M$_{\odot}$. While with our study to exclude any peculiar dynamical history for all these clusters, we at the same time seem to give
further supports support to the hypothesis that whether or not a cluster hosts multiple-stellar populations primarily depends from the mass and stellar density~\citep[see e.g.][]{bastian18}. However, abundance ratios for individual stars in the Halo GCs are urgently needed to assess the
presence of multiple stellar populations in these clusters and ultimately understand the relationship, investigated by recent studies~\citep[see e.g.][]{massari19, Malhan2021}, between remote GCs and other Halo substructures (i.e., luminous and ultra-faint dwarf spheroidal galaxies).



\begin{acknowledgements}
We thank the referee Christian Moni Bidin for their time and a helpful
report that has improved this paper. MC, ED acknowledge financial support 
from the project {\it Light-on-Dark} granted by MIUR through
PRIN2017-2017K7REXT. 
Based on observations made with the NASA/ESA Hubble Space Telescope, obtained from the data archive at the Space Telescope Science Institute. STScI is operated by the Association of Universities for Research in Astronomy, Inc. under NASA contract NAS 5-26555. This work has made use of data from the European Space Agency (ESA) mission
{\it Gaia} (\url{https://www.cosmos.esa.int/gaia}), processed by the {\it Gaia}
Data Processing and Analysis Consortium (DPAC,
\url{https://www.cosmos.esa.int/web/gaia/dpac/consortium}). Funding for the DPAC
has been provided by national institutions, in particular the institutions
participating in the {\it Gaia} Multilateral Agreement. This research has made use of the products of the Cosmic-Lab project funded by the European Research Council.
\end{acknowledgements}

%
%

\bibliographystyle{aa}
\bibliography{ref}

\end{document}